\documentclass[prd,nofootinbib,preprintnumbers,
]{revtex4}
\usepackage{graphicx,epsf,epsfig}
\usepackage{amsmath, amssymb, amsfonts, bbm}
\usepackage{amssymb}
\usepackage{tabls,multirow}

\def\vev#1{\left\langle #1\right\rangle}

\def\hbar{\hspace{0pt}\raisebox{1pt}{$-$} \hspace{-7pt} h}

\newcommand{\be}{\begin{equation}}
\newcommand{\ee}{\end{equation}}
\newcommand{\bd}{\begin{displaymath}}
\newcommand{\ed}{\end{displaymath}}
\newcommand{\bea}{\begin{eqnarray}}
\newcommand{\eea}{\end{eqnarray}}
\newcommand{\nn}{\nonumber}

\def\so10{$SO(10)$}

\begin{document}
\title{Higgs sector of the next-to-minimal renormalizable SUSY SO(10)}
\date{\today}
\author{Michal Malinsk\'{y}}\email{malinsky@phys.soton.ac.uk}
\affiliation{School of Physics and Astronomy, University of Southampton,
SO16 1BJ Southampton, United Kingdom}
\begin{abstract}
We study the Higgs potential of the next-to-minimal renormalizable SUSY SO(10) GUT with $\bf 120_{\rm H}$ Higgs representation on top of the ``standard'' minimal model Higgs sector spanning over $\bf 10_{\rm H}$,  $\bf \overline{126}_{\rm H}\oplus 126_{\rm H}$, $\bf 210_{\rm H}$. All the GUT-scale Higgs sector mass matrices for the 592 Higgs states of the model are written down in  detail with all the conventions fully specified. The consistency of the results is checked by the decoupling of ${\bf 120}_{\rm H}$ and independently by the analysis of the relevant Goldstone modes. The matching of the Yukawa sector sum-rules driving the matter fermion masses and mixing at the level of the effective theory is described thoroughly. 
\end{abstract}
\maketitle
\section{Introduction}
The idea of grand unification \cite{Georgi:1974sy} is one of the simplest and very powerful strategies of extending the Standard Model (SM) of the elementary particle interactions and the grand-unified theories (GUTs), though most of them formulated already in 1970's, receive still a lot of attention of the high energy physics community. Apart from accounting for all the Standard Model phenomena within a unified picture (accommodating the plethora of the  SM multiplets into just a few irreducible representation of the unified gauge group $G\supset SU(3)_{c}\otimes SU(2)_{L}\otimes U(1)_{Y}$) the GUTs typically provide answers to the questions the Standard theory is incapable to answer, like e.g. the hypercharge quantization or the hierarchy of the gauge (and some of the Yukawa) couplings at the electroweak scale. On top of that, GUTs generally predict distinctive new phenomena like proton decay, nucleon-antinucleon oscillations, monopole production in the early stage of the universe evolution etc., which makes these frameworks potentially testable and thus physical \cite{Popper}.

\vskip 2mm
One of the most popular grand-unified frameworks is based on the $SO(10)$ gauge group \cite{Fritzsch:1974nn} which is assumed to govern physics at very high scales (typically above $10^{16}$ GeV) providing for the the Standard Model dynamics once the GUT symmetries get spontaneously broken. Perhaps the main virtue of the $SO(10)$ models is that all the SM matter fermions of each generation resides in a 16-dimensional spinorial multiplet ${\bf 16}_{\rm F}$, which, however, contains the SM singlet playing the role of the right-handed neutrino, it is then very natural that small neutrino masses \cite{Strumia:2006db} are generated by means of some variant of the seesaw mechanism\cite{seesaw1}. Furthermore, as a rank-5 group, $SO(10)$ admits several different symmetry breaking patterns (see e.g. \cite{SO10breakingpattern,SO10breakingpattern2,Aulakh:2000sn,Barr:1981qv}), which is a feature that renders these theories slightly more flexible than the other popular scenarios based on the $SU(5)$ gauge symmetry \cite{Dimopoulos:1981zb}, alleviating to some degree the problems with proton decay (see e.g. \cite{SO10protondecay,SO10protondecay2}) generically emerging in the supersymmetric (SUSY) $SU(5)$ models.

\vskip 2mm
Recently, a lot of attention has been attracted to the class of SUSY $SO(10)$ models \cite{MinimalSO10} with large Higgs representations, in particular those with a 5-index antisymmetric tensor in the Higgs sector (decomposing under parity into ${\bf 126}_{\rm H}\oplus {\bf \overline{126}}_{\rm H}$) which (together with the 'traditional' ${\bf 10}_{\rm H}$) provides an option to construct a tightly constrained, yet potentially realistic, effective SM Yukawa sector at the renormalizable level \cite{MinimalSO10,Aulakh:2003kg}. The relevant superpotential $W_{Y}={\bf 16}_{\rm F}f_{10}{\bf 16}_{\rm F}{\bf 10}_{\rm H}+{\bf 16}_{\rm F}f_{\overline{126}}{\bf 16}_{\rm F}{\bf \overline{126}}_{\rm H}$ contains only a pair of unknown (symmetric) Yukawa matrices which subsequently give rise to all the SM Yukawa couplings in terms of simple linear combinations of $f_{10}$ and $f_{\overline{126}}$. Since, on top of that, the Majorana masses of neutrinos are governed by the same structure ($f_{\overline{126}}$), this class of models features a high degree of predictivity for the SM flavour, in particular the quark and lepton masses and mixings \cite{MinimalSO10Yukawastudies}. 

Subsequently, the simplest variants of this framework have been scrutinized in great detail \cite{MinimalSO10Yukawastudies,MinimalSO10examination}. It was even argued that the model with ${\bf 10}_{\rm H}\oplus{\bf 126}_{\rm H}\oplus {\bf \overline{126}}_{\rm H}\oplus {\bf 210}_{\rm H}$ in the Higgs sector and the three matter families residing in three copies of ${\bf 16}_{\rm F}$ of $SO(10)$ is indeed the minimal potentially realistic supersymmetric GUT (MSGUT) with only 26 (real) parameters driving the MSSM physics (up to the soft SUSY-breaking sector) \cite{Aulakh:2003kg} and an automate $R$-parity as a direct consequence of the $B-L$ symmetry breakdown by means of ${\bf 126}_{\rm H}\oplus {\bf \overline{126}}_{\rm H}$ \cite{Rparity}. Moreover, it was observed that in case of a dominant triplet contribution in the seesaw formula for the neutrino masses, there is a natural case for the maximality of the atmospheric mixing in the lepton sector due to the GUT-scale $b$- and $\tau$- Yukawa coupling convergence \cite{b-taularge23}. On top of that, a generic tendency for a somewhat larger 13 mixing in the lepton sector ($|U_{e3}|\gtrsim 0.1$) has been observed, providing a clear experimental signal within the reach of the near-future facilities \cite{Ue3measurements}. This, of course,  triggered an enormous boost to the field with tens of papers concerning various aspects of the model, like e.g. proton decay \cite{SO10protondecay}, effective Yukawa sector \cite{MinimalSO10Yukawastudies}, gauge coupling unification \cite{Aulakh:2000sn} etc. 

However, the real breakthrough came in 2004 with the study by Bajc, Melfo, Senjanovi\'{c} and Vissani \cite{Bajc:2004xe} (preceded by \cite{Lee:1993rr} accounting for some partial results) providing a complete analysis of the Higgs potential of the minimal model, deriving in particular the mass matrices of all the SM components of the 472-dimensional Higgs sector in terms of the parameters of the underlying $SO(10)$ lagrangian. 
With the extra information at hand it was possible to improve the existing analyses to such a degree a tension between the minimal theory and observation has been revealed \cite{Aulakh:2005bd,absoluteneutrinomassscale,Aulakh:2006vi}. It was argued that the need for a relatively low singlet Majorana neutrino mass scale (of the order of $10^{13-14}$ GeV) in order to satisfy the absolute neutrino mass constraints is indeed in clash with the fits of the effective Yukawa sector sum-rules. Basically, the trouble stems from the dual role of the ${\bf 126}_{\rm H}$ multiplet in the Yukawa sector: the need for the $B-L$ breakdown close to the GUT-scale ($M_{G}$) which is mandatory \cite{Bajc:2004xe} in order to prevent the Higgs sector thresholds from spoiling gauge-coupling unification requires a relatively small Yukawa coupling of ${\bf 126}_{\rm H}$ which is typically at odds with the second generation SM matter hierarchy.    

\vskip 2mm
Perhaps the most conservative extension of the minimal model\footnote{We shall not comment at all on the potentially predictive $SO(10)$ scenarios exploiting an alternative setting of the Higgs sector based on $16\oplus \overline{16}$ rather than $126\oplus \overline{126}$; an interested reader can find a representative sample of references in papers \cite{Babu:1993we,vectormatter}.} that could alleviate this tension, yet retaining a significant amount of its predictive power, consists in invoking an extra 120-dimensional three-index antisymmetric tensor in the Higgs sector \cite{NMSGUT, adding120,Aulakh:2007ir}. Such a scenario, sometimes called 'the next-to minimal SUSY $SO(10)$' (or 'new minimal  SUSY $SO(10)$', NMSGUT \cite{Aulakh:2007ir}) has the appealing feature of accounting for all three possible types of renormalizable couplings the matter bilinear ${\bf 16}_{\rm F}.{\bf 16}_{\rm F}$ can potentially develop (recall that  ${\bf 16}\otimes{\bf 16}={\bf 10}\oplus {\bf 126}\oplus {\bf 120}$) and the price to be paid for the extra multiplet can be acceptable - the Yukawa coupling of ${\bf 120}_{\rm H}$ is antisymmetric and thus the minimal model Yukawa sector predictivity is challenged in a minimal way. With an extra set of couplings in the game, the effective SM Yukawa sum-rules are altered, which provides an option to account for the second generation hierarchies by means of the Yukawa of ${\bf 120}_{\rm H}$ while keeping $f_{\overline{126}}$ relatively free to deal with the MSGUT neutrino mass scale problem \cite{Aulakh:2007ir}. Since the full description of the SM quark and lepton masses based on the interplay of the ${\bf {10}}_{\rm H}$ and ${\bf {120}}_{\rm H}$ only seems troublesome \cite{Lavoura:2006dn}, the subleading effects from ${\bf \overline{126}}_{\rm H}$ deferred almost entirely for the sake of the neutrino sector can be just the right cure to reconcile the first and second generation structures with the observation, which makes this extension of the minimal model particularly interesting and worth further scrutiny. 

However, due to the further relaxation of the effective sum-rules via an extra Yukawa in the model the old-fashioned effective analyses of the Yukawa sum-rules became indecisive in the NMSGUT case (though some interesting results in this direction have been obtained under various extra assumptions like e.g. perturbatively small ${\bf 120}_{\rm H}$-effects \cite{adding120}, in a setting with decoupled ${\bf \overline{126}}_{\rm H}$ \cite{Lavoura:2006dn} or imposing furhter constraints on the Yukawa couplings, see e.g. \cite{Grimus:2006rk}). Any extra information about the GUT-scale Higgs  spectra and mixings (as almost the only means of constraining the parametric space of the model) became a crucial ingredient of any quantitative analysis of the model.        
\vskip 2mm
The main scope of this work is to study the Higgs sector of the NMSGUT case (with the total of ${\bf 10}_{\rm H}\oplus{\bf 126}_{\rm H}\oplus {\bf \overline{126}}_{\rm H}\oplus {\bf 210}_{\rm H}\oplus {\bf 120}_{\rm H}$ in the Higgs sector) as thoroughly as possible, i.e. calculate all the Higgs sector mass matrices (encoding the information about the GUT-scale Higgs spectra and mixings which is the key to a reliable analysis of the effective Yukawa sector). We shall pass through all the SM submultiplets of the 592-dimensional Higgs sector of NMSGUT providing the relevant mass matrices for each sector in the form that emerges right after the breakdown of the GUT symmetry and comment on some of the features of the different multiplets. Last, but not least, in the limit when ${\bf 120}_{\rm H}$ decouples from the GUT-scale physics, our results provide an independent cross-check of the MSGUT formulae given in \cite{Bajc:2004xe}.

\vskip 2mm
The study is organized as follows: after recapitulating in brief some of the salient features of the Higgs sector of the minimal model we write down an upgraded form of the Higgs superpotential relevant for the NMSGUT case (including the extra ${\bf 120}_{\rm H}$) in Section \ref{superpotential} and argue (Section \ref{breakdown}) that despite from the extra 5 terms therein the vacuum structure of the extended model is the same like that of the MSGUT. This allows us to inherit the minimal model parametrization of the vacuum manifold from \cite{Bajc:2004xe} and write down all the mass matrices in Section \ref{sect-massmatrices} in the 'familiar' notation. Each of the mass matrices is augmented with a complete information on the states it is spanned over and on the phase convention employed upon its derivation so that the study is maximally self-contained and verifiable. Section \ref{goldstones} is then devoted to the identification of the Goldstone modes in the relevant sectors, which provides a further consistency check of our results. Focusing on the Yukawa sector of the MSSM in Section \ref{sect-yukawasector}, we provide a detailed description of the matching between the effective theory and the underlying $SO(10)$ model. Most of the technicalities (namely notation issues \& comments on conventions) are deferred to an Appendix. 
\section{The NMSGUT Higgs sector superpotential\label{superpotential}}
Using the ``group theory notation'' the Higgs part of the superpotential of the minimal SUSY $SO(10)$ model reads:
\bea
W_{\rm H}^{\mathrm{min}} & = & M_{10} {\bf 10_{\rm H}}.{\bf 10_{\rm H}}+M_{126} {\bf \overline{126}_{\rm H}. 126_{\rm H}} +M_{210} {\bf 210_{\rm H}}.{\bf 210_{\rm H}}  +\nn \\
& + &\lambda\, {\bf 210_{\rm H}}.{\bf 210_{\rm H}}.{\bf 210_{\rm H}}+\eta\,{\bf 210_{\rm H}}.{\bf \overline{126}_{\rm H}}.{\bf 126_{\rm H}} +  \alpha\,{\bf  210_{\rm H}}.{\bf 10_{\rm H}}.{\bf 126_{\rm H}}+\overline{\alpha}\,{\bf  210_{\rm H}}.{\bf 10_{\rm H}}.{\bf \overline{126}_{\rm H}}\label{Wmin}
\eea
which is the piece that has been studied in detail in \cite{Bajc:2004xe}. 
In terms of the $SO(10)$ tensorial components:
\be
H_{i}\equiv {\bf 10_{\rm H}}\;,\quad  
\Phi_{ijkl}\equiv {\bf 210_{\rm H}}\;,\quad  
\Sigma_{ijklm}\equiv {\bf 126_{\rm H}}\;,\quad  
\overline{\Sigma}_{ijklm}\equiv {\bf \overline{126}_{\rm H}}\;,
\label{components}
\ee
$W_{\rm H}^{\mathrm{min}}$ can be transcribed as:
\bea
W_{\rm H}^{\mathrm{min}} & = & m_{\rm H}{H_{i}}H_{i}+\frac{1}{5!}m_{\Sigma}{\Sigma}_{ijklm}\overline\Sigma_{ijklm}+\frac{1}{4!}m_{\Phi}\Phi_{ijkl}\Phi_{ijkl}+\nn\\
&+& \frac{1}{4!}\lambda\,\Phi_{ijkl}\Phi_{klmn}\Phi_{mnij}+\frac{1}{4!}\eta\,\Phi_{ijkl}\Sigma_{ijmno}\overline{\Sigma}_{klmno}+\frac{1}{4!}\alpha\,\Phi_{ijkl}H_{m}\Sigma_{ijklm}+\frac{1}{4!}\overline{\alpha}\,\Phi_{ijkl}H_{m}\overline{\Sigma}_{ijklm}\label{Wmin2}
\eea
where the notation of \cite{Bajc:2004xe} has been fully inherited and summation over all repeating indices is understood. There is in total 7 parameters in $W_{\rm H}^{\mathrm{min}}$: $\{m_{\rm H}, m_{\Sigma}, m_{\Phi}, \alpha,\overline{\alpha},\eta,\lambda\}$. However, 1 of them  ($m_{\Sigma}$) can be traded for a dimensionless parameter called $x$ from the requirements on the GUT symmetry breaking chain and another (conventionally $m_{\rm H}$) can be fixed by the need for a zero determinant of the $SU(2)_{L}$ doublet mass matrix to arrange the pair of light MSSM Higgs doublets.

\vskip 2mm
In NMSGUT, the extra 120-dimensional three index fully antisymmetric tensor is added to the Higgs sector giving rise to a new set of vertices with $\bf 120_{\rm H}$:
\bea
W_{\rm H}^{\bf 120}& = &  M_{120} {\bf 120_H}.{\bf 120_H}+ \nn \\
& + & \rho\,{\bf 210_{\rm H}}.{\bf 120_{\rm H}}.{\bf 120_{\rm H}}+ \gamma\,{\bf 10_{\rm H}}.{\bf 210_{\rm H}}.{\bf 120_{\rm H}}
 +  \beta\,{\bf  210_{\rm H}}.{\bf 120_{\rm H}}.{\bf 126_{\rm H}}+\overline{\beta}\,{\bf  210_{\rm H}}.{\bf 120_{\rm H}}.{\bf \overline{126}_{\rm H}}
\label{superpot}
\eea
Using
$\Psi_{ijk}\equiv {\bf 120}_{\rm H}$
for the components of ${\bf 120_{\rm H}}$ we can rewrite $W_{\rm H}^{\bf 120}$ in the form:
\bea
W_{\rm H}^{\bf 120}& = & \frac{1}{3!}m_{\Psi}\Psi_{ijk}\Psi_{ijk}+\nn \\
& + & \frac{1}{(2!)^{2}}\rho\,\Phi_{ijkl}\Psi_{ijm}\Psi_{klm}+\frac{1}{3!}\gamma\,H_{i}\Phi_{ijkl}\Psi_{jkl}+\frac{1}{3!}\beta\,\Phi_{ijkl}\Psi_{lmn}\Sigma_{ijkmn}+\frac{1}{3!}\overline{\beta}\,\Phi_{ijkl}\Psi_{lmn}\overline{\Sigma}_{ijkmn}\label{W120}
\eea
This piece of the total NMSGUT superpotential $W_{\rm H}=W_{\rm H}^{\mathrm{min}}+W_{\rm H}^{\bf 120}$
brings in 5 extra parameters: $\{m_{\Psi},\beta, \overline{\beta},\gamma,\rho\}$. All together, there is 11 relevant parameters in the next-to-minimal setting, namely $\{x, m_{\Phi}, m_{\Psi}, \alpha,\overline{\alpha},\eta,\lambda,\beta, \overline{\beta},\gamma,\rho\}$ that we shall further refer to as ``microscopic''.

\vskip 2mm
In what follows, we shall mostly use the notation and conventions defined in the Appendix A.1 of \cite{Bajc:2004xe}. However, since the embedding of the SM states into the $SO(10)$ multiplets is defined only up to global phases, there will be some differences in the choice of phase convention for the basis states the mass matrices are spanned over. For sake of completeness, we will comment on this issue wherever appropriate. Further information on accounting for some of the convention differences is given in Appendix \ref{decompositions}.   
\section{GUT symmetry breakdown\label{breakdown}}
The minimal model Higgs multiplets ${\bf 10}_{\rm H}$, ${\bf \overline{126}}_{\rm H}\oplus {\bf 126}_{\rm H}$ and ${\bf 210}_{\rm H}$ decomposing under the Pati-Salam subgroup $SU(4)\otimes SU(2)_{L}\otimes SU(2)_{R}$ of $SO(10)$ as 
\bea
{\bf 10} &=& (1,2,2)\oplus (6,1,1)\nn \\
{\bf \overline{126}} &=& (6,1,1)\oplus (\overline{10},3,1)\oplus({10},1,3)\oplus (15,2,2) \nn\\
{\bf {126}} &=& (6,1,1)\oplus (10,3,1)\oplus(\overline{10},1,3)\oplus (15,2,2) \\
{\bf 210} &=& (1,1,1)\oplus (15,1,1)\oplus (6,2,2)\oplus (15,3,1)\oplus (15,1,3)\oplus (10,2,2)\oplus(\overline{10},2,2)\nn
\eea
accommodate in total 5 complete singlets under the Standard Model gauge group $SU(3)_{c}\otimes SU(2)_{L}\otimes U(1)_{Y}$ residing in the $({10},1,3)_{\overline{126}}$, $(\overline{10},1,3)_{{126}}$, $(1,1,1)_{{210}}$, $(15,1,3)_{{210}}$ and $(15,1,1)_{{210}}$ components of the relevant $SO(10)$ tensors. This can be seen from the decompositions of the Pati-Salam multiplets into the $SU(3)_{c}\otimes U(1)_{B-L}$ states
\bea
{\bf 4}&\equiv&(3,+\tfrac{1}{3}) \oplus (1,-1),\qquad\qquad\qquad\;\, 
{\bf  6}\;=\;(3,-\tfrac{2}{3}) \oplus (\bar 3,+\tfrac{2}{3})\nn\\
{\bf  10}&=&(6,+\tfrac{2}{3})\oplus (3,-\tfrac{2}{3}) \oplus (1,-2)\qquad
{\bf  15}\;=\;(8,0)\oplus (1,0)\oplus (3,+\tfrac{4}{3}) \oplus (\bar 3,-\tfrac{4}{3})\label{PSmultiplets}
\eea
provided $Q=T^{L}_{3}+Y$, i.e. $Y=T^{R}_{3}+\tfrac{1}{2}(B-L)$ is being used as a  convention for the Gell-Mann Nishijima formula defining the weak hypercharge normalization.

In what follows we shall (in accordance with \cite{Bajc:2004xe})  denote the vacuum expectation values (VEVs) of the relevant SM singlets in these multiplets as:
\be\label{VEVs}
\vev{({10},1,3)_{\overline{126}}}\equiv \overline{\sigma},\;\;
\vev{(\overline{10},1,3)_{{126}}}\equiv \sigma,\;\;
\vev{(1,1,1)_{{210}}}\equiv p,\;\;
\vev{(15,1,3)_{{210}}}\equiv \omega \;\;\text{and}\;\;
\vev{(15,1,1)_{{210}}}\equiv a.
\ee 
 
As it was shown in \cite{Bajc:2004xe} there are 7 distinct SUSY preserving minima of the minimal model Higgs potential corresponding to $SO(10)$, $SU(5)\otimes U(1)$, flipped $SU(5)\otimes U(1)$, $SU(4)\otimes SU(2)_{L}\otimes SU(2)_{R}$, $SU(3)_{c}\otimes SU(2)_{L}\otimes SU(2)_{R}\otimes U(1)_{B-L}$, $SU(3)_{c}\otimes SU(2)_{L}\otimes U(1)_{R}\otimes U(1)_{B-L}$ and $SU(3)_{c}\otimes SU(2)_{L}\otimes U(1)_{Y}$ vacua respectively and there is a single dimensionless parameter called $x$ in \cite{Bajc:2004xe} governing all the relevant breaking chains.
The Standard Model vacuum manifold then corresponds to those values of $x$ which generate ``generic'' patterns amongst the VEVs of the Standard Model singlet VEVs
\be\label{SMvacuumparameters}
\omega\equiv -\frac{m_{\Phi}}{\lambda}x,\qquad
p\equiv \frac{m_{\Phi}}{\lambda}\frac{x(5x^{2}-1)}{(x-1)^{2}},\qquad
a\equiv \frac{m_{\Phi}}{\lambda}\frac{x^{2}+2x-1}{x-1},\qquad
m_{\Sigma}=-\frac{\eta}{\lambda}m_{\Phi}\frac{8 x^3-15 x^2+14 x-3}{(x-1)^2}
\ee
and 
\be\label{sses}
\sigma\overline\sigma=\frac{2m_{\Phi}^{2}}{\eta\lambda}\frac{x(x^{2}+1)(1-3x)}{(x-1)^{2}}\text{\;\;\;or\;\;\;}s\overline s=\frac{2x(x^{2}+1)(1-3x)}{(x-1)^{2}}\text{\;\;\;with\;\;\;} s\equiv \frac{m_{\Phi}}{\sqrt{\lambda\eta}}\sigma \text{\;\;\;etc.}
\ee
in which case $SO(10)$ is broken down to  $SU(3)_{c}\otimes SU(2)_{L}\otimes U(1)_{Y}$ essentially in one step while the intermediate symmetries can be obtained if $x$ is tuned to receive one of the particular values identified in \cite{Bajc:2004xe}. Let us remind the reader that the GUT-scale $D$-flatness implies $\sigma=\bar\sigma$ up to an overall phase. 

In view of the complicated structure of the minimal model vacuum it is rather welcome that the extra ${\bf 120}_{\rm H}$ decomposing under the Pati-Salam subgroup as
\bea
{\bf {120}} &=& (1,2,2)\oplus (10,1,1)\oplus(\overline{10},1,1)\oplus (6,3,1)\oplus(6,1,3)\oplus(15,2,2) 
\eea
does not bring in any extra SM singlets because the only colour singlet of $\bf 10$ of Pati-Salam has a non-zero $B-L$ charge, c.f. formulae (\ref{PSmultiplets}). This means that the vacuum structure of the minimal model remains unaffected even with an extra ${\bf 120}_{\rm H}$ in the game, and in particular the parametrization (\ref{SMvacuumparameters}) can be entirely inherited for sake of the discussion of the extended model. This feature saves us a lot of tedium indeed.

\section{Higgs sector mass matrices\label{sect-massmatrices}}
In this section we shall provide the mass matrices for the (fermionic components of the) Higgs multiplets after the SO(10) breakdown to the Standard Model $SU(3)_{c}\otimes SU(2)_{L}\otimes U(1)_{Y}$. In order to simplify the situation considerably we shall, as usual, focus on the fermionic (i.e. higgsino) components of the superfields and invoke SUSY (which remains unbroken at the GUT scale) to transfer the results of interest (i.e. the eigenvalues and mixings) into the bosonic sector. The main advantage of this approach is that the higgsino mass matrices can be obtained directly from the VEVs of the double superpotential derivatives and there is no need to pass through the lengthy computation of all the $F$-terms and $D$-terms providing the building blocks of the SUSY scalar potential. 

For sake of illustration let us remark that there is in total 13,321,010 terms in the sums in (\ref{Wmin2}) and (\ref{W120}) (out of which 2,111 thousand terms come from the new piece $W_{\rm H}^{\bf 120}$), but fortunately 'only' 1,190,170 of them are non-zero by antisymmetry of the tensors under consideration ($W_{\rm H}^{\bf 120}$ then accounts for 338,400 out of this number). Thus, perhaps the only reasonable strategy of handling all these contributions is to work with the antisymmetrized combinations rather than with the very components of the antisymmetric tensors (this is why all the results in \cite{Bajc:2004xe} are given just in terms of such structures), c.f. Appendix \ref{decompositions}.
\subsection{General remarks}
At the Standard Model level, all states with the same quantum numbers with respect to the SM Cartan operators $T^{c}_{3}$,  $T^{c}_{8}$,  $T^{L}_{3}$, $Y$ corresponding to the same Casimir eigenvalues of $C^{c}\equiv \sum T^{c2}_{i}$ and $C^{L}\equiv \sum T^{L2}_{i}$ can mix and a characteristic pattern of mass matrices emerges.
In what follows we shall pass through the whole plethora of the Higgs sector states and write down the corresponding mass (fermionic) matrix for each subspace corresponding to a set of fixed values of ($C^{c}$, $T^{c}_{3}$,  $T^{c}_{8}$; $C^{L}$,  $T^{L}_{3}$; $Y$) or, equivalently
,   ($C^{c}$, $T^{c}_{3}$,  $T^{c}_{8}$; $B-L$, $C^{L}$,  $T^{L}_{3}$; $C^{R}$,$T^{R}_{3}$)  choosing a single representative configuration of the Cartan eigenvalues for each value of the relevant Casimir, for example $T^{c}_{3}=0$,  $T^{c}_{8}=0$ state for the $SU(3)$ octets (corresponding to $C^{c}=\tfrac{10}{3}$) or e.g. $T^{L}_{3}=-\tfrac{1}{2}$ for the $SU(2)_{L}$ doublets (with $C^{L}=\tfrac{3}{4}$) etc. The mass matrices for all the other components with the same $C^{c}$, $C^{L}$, $Y$ corresponding to just different $T^{c}_{3}$,  $T^{c}_{8}$ and/or $T^{L}_{3}$ can be (if desired) obtained in a straightforward manner by the relevant $SU(3)_{c}$ and/or $SU(2)_{L}$ transformations. 

In every sector (i.e. for every combination of $C^{c}$, $C^{L}$, $Y$), we shall first specify the choice of the representative (i.e. the values of $T^{c}_{3}$,  $T^{c}_{8}$, $T^{L}_{3}$) and comment on the dimensionality and origin of the relevant contributions. Every subsequent mass matrix shall be equipped with a table of states corresponding to its rows (denoted generically by $R_{i}$) and columns ($C_{i}$), unless the representation is real, in which case a single symbol $S_{i}$ shall be used. For each $C_{i}$ and $R_{i}$ (or $S_{i}$) we shall also display a chunk of the map of the SM components of ${\bf 10}_{\rm H}$, ${\bf 126}_{\rm H}$, ${\bf \overline{126}}_{\rm H}$, ${\bf 210}_{\rm H}$ and ${\bf 120}_{\rm H}$ (i.e. the submultiplets with definite SM quantum numbers) onto the defining basis states $H_{i}$, $\Sigma_{ijklm}$, $\overline\Sigma_{ijklm}$, $\Phi_{ijkl}$ and $\Psi_{ijk}$ (typically we present only the ``lowest'' relevant permutation of indices and defer an interested reader to Appendix \ref{decompositions} or to \cite{Bajc:2004xe} for further details) in order to provide an information about the phase convention used in derivation of the mass matrix under consideration. (Note that for sake of simplicity we always choose our phase convention in such a way there are no pending imaginary units in the mass matrices.) For sake of a simple bookkeeping the top-left box of each table shall indicate the full dimensionality of the sector under consideration.
\subsection{Mass matrices - colour singlets}
There are in total 40 colour singlet states in the Higgs sector of the minimal model. With an extra ${\bf 120}_{\rm H}$ this number is increased up to 50. In what follows we shall pass through the three categories corresponding to $SU(2)_{L}$ singlets, doublets and triplets respectively. 
\subsubsection{$C^{c}=0,  C^{L}=0$ - $SU(3)_{c}\otimes SU(2)_{L}$ singlets}
There are altogether 13  different $SU(3)_{c}\otimes SU(2)_{L}$ singlet states in the NMSGUT breaking into three different hypercharge sectors with $Y=0$, $\pm 1$ and $\pm 2$ respectively.

As in the minimal model case, the $Y=0$ sector of the  NMSGUT consists of 5 entirely SM-neutral Higgses (and thus the representation is real and the mass matrix symmetric) because there is no full SM singlet in ${\bf 120}_{\rm H}$. Note also that these are the only fields that can receive GUT-scale VEVs triggering the $SO(10)$ breakdown.
The $Y=\pm 1$ sector of the NMSGUT differs from the minimal model one due to an extra state in ${\bf 120}_{\rm H}$ and the relevant $3\times 3$ mass matrix is hermitean accounting for 6 degrees of freedom in total. Notice that these are exactly the quantum numbers of the $SU(2)_{R}$ gauge bosons ($W_{R}$) that should be made superheavy via the standard Higgs mechanism once the $SU(2)_{R}$ symmetry gets broken. In order to be able to achieve that, there must be a Goldstone boson in this sector corresponding to a zero in the spectrum of the relevant mass matrix. For further details the reader is deferred to Section \ref{goldstones}.
Finally, the 2 components with $Y=\pm 2$ are identical to the minimal model ones (coming from ${\bf \overline{126}}_{\rm H}{\bf \oplus 126}_{\rm H}$) as well as the corresponding mass matrix.

The relevant mass matrices and the corresponding physical basis vectors (and the phase convention) these matrices are spanned on follow:
\vskip 5mm
\paragraph{\bf Sector $(1,1,0)$:}
\be
\left(
\begin{array}{ccccc}
 0 &m_{\Sigma}+\eta  (3 a+p-6 \omega ) & -\eta  \overline{\sigma} & -\sqrt{3} \eta  \overline{\sigma} & \sqrt{6} \eta  \overline{\sigma} \\
m_{\Sigma}+\eta  (3 a+p-6 \omega ) & 0 & -\eta  \sigma  & -\sqrt{3} \eta  \sigma  & \sqrt{6} \eta 
   \sigma  \\
 -\eta  \overline{\sigma} & -\eta  \sigma  & 2 m_{\Phi} & 0 & 2 \sqrt{6} \lambda  \omega  \\
 -\sqrt{3} \eta  \overline{\sigma} & -\sqrt{3} \eta  \sigma  & 0 & 2 (m_{\Phi}+2 a \lambda ) & 4
   \sqrt{2} \lambda  \omega  \\
 \sqrt{6} \eta  \overline{\sigma} & \sqrt{6} \eta  \sigma  & 2 \sqrt{6} \lambda  \omega  & 4 \sqrt{2}
   \lambda  \omega  & 2 (m_{\Phi}+(2 a+p) \lambda )
\end{array}
\right)
\ee
\begin{center}
\begin{tabular}{|c|c|c|c|}
\hline 5 & Pati-Salam origin & ($C^{c}\!\!, T^{c}_{3},  T^{c}_{8}, B-L, C^{L}\!\!,T^{L}_{3}, C^{R}\!\!,  T^{R}_{3}$) &  phase convention\\
\hline
$S_{1}$ & $(\overline{10},1,3)_{126}$ & $(0,0,0,2,0,0,2,-1)_{126}$ & $-\Sigma[1,3,5,7,9]+\ldots$ \\
$S_{2}$ & $({10},1,3)_{\overline{126}}$ & $(0,0,0,-2,0,0,2,1)_{\overline{126}}$ & $-\overline{\Sigma}[1,3,5,7,9]+\ldots$\\
$S_{3}$ & $(1,1,1)_{210}$ & $(0,0,0,0,0,0,0,0)_{210}$ & $\Phi[1,2,3,4]$ \\
$S_{4}$ &  $(15,1,1)_{210}$& $(0,0,0,0,0,0,0,0)_{210}$ & $\Phi[5,6,7,8]+\ldots$ \\
$S_{5}$ &  $(15,1,3)_{210}$& $(0,0,0,0,0,0,2,0)_{210}$ & $\Phi[1,2,5,6]+\ldots$ \\
\hline \end{tabular}
\end{center}
\vskip 5mm
\paragraph{\bf Sector $(1,1,-1)\oplus (1,1,+1)$:}
\be
\left(
\begin{array}{lll}
 m_\Sigma+(3 a+p) \eta  & \sqrt{6} \eta  \sigma  & 6 \sqrt{2} \bar\beta \omega  \\
 \sqrt{6} \eta  \overline{\sigma} & 2 (m_\Phi+(2 a+p) \lambda ) & 2 \sqrt{3} \bar\beta
   \overline{\sigma} \\
 6 \sqrt{2} \beta  \omega  & 2 \sqrt{3} \beta  \sigma  & 2 (m_\Psi+3 a \rho )
\end{array}
\right)\ee
\begin{center}
\begin{tabular}{|c|c|c|c|}
\hline 6 & Pati-Salam origin & ($C^{c}\!\!, T^{c}_{3},  T^{c}_{8}, B-L, C^{L}\!\!,T^{L}_{3}, C^{R}\!\!,  T^{R}_{3}$) &  phase convention\\
\hline
$R_{1}$ & $({10},1,3)_{\overline{126}}$&  $(0,0,0,-2,0,0,2,0)_{\overline{126}}$ & $-i \overline{\Sigma}[1,2,5,7,9]+\ldots$ \\
$R_{2}$ & $(15,1,3)_{210}$ &  $(0,0,0,0,0,0,2,-1)_{210}$ & $-i \Phi[1,3,5,6]+\ldots$\\
$R_{3}$ & $({10},1,1)_{{120}}$ &  $(0,0,0,-2,0,0,0,0)_{120}$ & $-\Psi[5,7,9]+\ldots$\\
\hline
$C_{1}$ &  $(\overline{10},1,3)_{{126}}$ &  $(0,0,0,+2,0,0,2,0)_{126}$ &  $+i \Sigma[1,2,5,7,9]+\ldots$ \\
$C_{2}$ & $(15,1,3)_{210}$&  $(0,0,0,0,0,0,2,+1)_{210}$ &  $+i \Phi[1,3,5,6]+\ldots$\\
$C_{3}$ & $(\overline{10},1,1)_{{120}}$ &  $(0,0,0,+2,0,0,0,0)_{120}$ & $-\Psi[5,7,9]+\ldots$\\
\hline \end{tabular}
\end{center}
\vskip 5mm
\paragraph{\bf Sector $(1,1,-2)\oplus (1,1,+2)$:}
\be
 m_\Sigma+\eta  (3 a+p+6 \omega )
\ee
\begin{center}
\begin{tabular}{|c|c|c|c|}
\hline 2 & Pati-Salam origin & ($C^{c}\!\!, T^{c}_{3},  T^{c}_{8}, B-L, C^{L}\!\!,T^{L}_{3}, C^{R}\!\!,  T^{R}_{3}$) &  phase convention\\
\hline
$R_{1}$ & $({10},1,3)_{\overline{126}}$&  $(0,0,0,-2,0,0,2,-1)_{\overline{126}}$ & $-\overline{\Sigma}[1,3,5,7,9]+\ldots$ \\
\hline
$C_{1}$ & $(\overline{10},1,3)_{{126}}$ &  $(0,0,0,+2,0,0,2,+1)_{126}$ & $-\Sigma[1,3,5,7,9]+\ldots$ \\
\hline \end{tabular}
\end{center}
One can easily check that in all three cases our results agree with the relevant matrices in \cite{Bajc:2004xe} up to row/column reshuffling and the phase conventions. 
\subsubsection{$C^{c}=0,  C^{L}=\frac{3}{4}$ - singlets of $SU(3)_{c}$, doublets of $SU(2)_{L}$}
There are in total 14 different colourless doublets accommodating 28 independent components. This cathegory breaks down into 2 subspaces with hypercharges $Y=\pm\frac{1}{2}$ and  $\pm\frac{3}{2}$. 

The $Y=\pm \frac{1}{2}$ sector corresponds to the MSSM-like Higgs doublets driving the Dirac masses of the matter sector. Due to the two extra Pati-Salam-bidoublets residing in ${\bf 120}_{\rm H}$,  the original MSGUT $4\times 4$ mass matrix for each charge is in NMSGUT upgraded to $6\times 6$, accounting for 12 doublets (i.e. 24 states in total).  We give the relevant mass matrix for the  $Q=0$ sector only; the same for $Q=\pm 1$ states is readily obtained from the $SU(2)_{L}$ symmetry considerations. (Since this is the key to any Yukawa sector analysis, a detailed study of this sector is subject of a dedicated Section \ref{sect-yukawasector}.) The remaining 4 states (i.e. 2 isodoublets) with $Y=\pm \frac{3}{2}$ are the same like in the minimal model.
\vskip 5mm
\paragraph{\bf Sector $(1,2,-\frac{1}{2})\oplus (1,2,+\frac{1}{2})$, $Q=0$ states:}
\be
\label{doubletmassmatrix}
\begin{pmatrix}
2 m_{\rm H} & 
-\sqrt{\frac{3}{2}}\overline\alpha(a-\omega)&
\sqrt{\frac{3}{2}}\alpha(a+\omega)&
-\overline{\alpha}\overline{\sigma}&
p\gamma &
\sqrt{3}\gamma \omega\\
-\sqrt{\frac{3}{2}}\alpha(a-\omega)&
m_{\Sigma}+2\eta(a-\omega)&
0&
\sqrt{6}\eta \overline{\sigma}&
\sqrt{6}\beta \omega&
\sqrt{2}\beta (p-2\omega)\\
\sqrt{\frac{3}{2}}\overline\alpha(a+\omega)&
0&
m_{\Sigma}+2\eta(a+\omega)&
0&
\sqrt{6}\overline{\beta} \omega&
\sqrt{2}\overline{\beta} (p+2\omega)\\
-\alpha\sigma&
\sqrt{6}\eta \sigma&
0&
2[m_{\Phi}+3\lambda(a-\omega)]&
2\beta\sigma&
-2\sqrt{3}\beta\sigma\\
p\gamma &
\sqrt{6}\overline{\beta} \omega&
\sqrt{6}\beta \omega&
2\overline{\beta}\overline{\sigma}&
2m_{\Psi}&
2\sqrt{3}\rho\omega \\
\sqrt{3}\gamma \omega&
\sqrt{2}\overline{\beta} (p-2\omega)&
\sqrt{2}\beta (p+2\omega)&
-2\sqrt{3}\overline{\beta}\overline{\sigma}&
2\sqrt{3}\rho\omega&
2m_{\Psi}+4a\rho
\end{pmatrix}.
\ee
\begin{center}
\begin{tabular}{|c|c|c|c|}
\hline 24 & Pati-Salam origin & ($C^{c}\!\!, T^{c}_{3},  T^{c}_{8}, B-L, C^{L}\!\!,T^{L}_{3}, C^{R}\!\!,  T^{R}_{3}$) &  phase convention\\
\hline
$R_{1}$ & $(1,2,2)_{10}$ &  $(0,0,0,0,\tfrac{3}{4},+\tfrac{1}{2},\tfrac{3}{4},-\tfrac{1}{2})_{10}$ & $-i H[3]+\ldots$ \\
$R_{2}$ & $(15,2,2)_{126}$ &  $(0,0,0,0,\tfrac{3}{4},+\tfrac{1}{2},\tfrac{3}{4},-\tfrac{1}{2})_{{126}}$ & $-i \Sigma [1,2,3,5,6]+\ldots$ \\
$R_{3}$ & $(15,2,2)_{\overline{126}}$ &  $(0,0,0,0,\tfrac{3}{4},+\tfrac{1}{2},\tfrac{3}{4},-\tfrac{1}{2})_{\overline{126}}$ & $-i \overline{\Sigma} [1,2,3,5,6]+\ldots$ \\
$R_{4}$ & $({10},2,2)_{210}$&  $(0,0,0,-2,\tfrac{3}{4},+\tfrac{1}{2},\tfrac{3}{4},+\tfrac{1}{2})_{210}$ & $-i\Phi[1,5,7,9]+\ldots$ \\
$R_{5}$ &  $(1,2,2)_{120}$ &  $(0,0,0,0,\tfrac{3}{4},+\tfrac{1}{2},\tfrac{3}{4},-\tfrac{1}{2})_{120}$ & $-\Psi[1,2,3] +\ldots$ \\
$R_{6}$ &  $(15,2,2)_{120}$ &  $(0,0,0,0,\tfrac{3}{4},+\tfrac{1}{2},\tfrac{3}{4},-\tfrac{1}{2})_{120}$ & $-\Psi[3,5,6] +\ldots$ \\
\hline
$C_{1}$ & $(1,2,2)_{10}$&  $(0,0,0,0,\tfrac{3}{4},-\tfrac{1}{2},\tfrac{3}{4},+\tfrac{1}{2})_{10}$ & $ i H[3] +\ldots$ \\
$C_{2}$ & $(15,2,2)_{\overline{126}}$&  $(0,0,0,0,\tfrac{3}{4},-\tfrac{1}{2},\tfrac{3}{4},+\tfrac{1}{2})_{\overline{126}}$ & $i \overline{\Sigma} [1,2,3,5,6]+\ldots$ \\
$C_{3}$ & $(15,2,2)_{126}$&  $(0,0,0,0,\tfrac{3}{4},-\tfrac{1}{2},\tfrac{3}{4},+\tfrac{1}{2})_{{126}}$ & $ i \Sigma [1,2,3,5,6]+\ldots$ \\
$C_{4}$ & $(\overline{10},2,2)_{210}$ &  $(0,0,0,+2,\tfrac{3}{4},-\tfrac{1}{2},\tfrac{3}{4},-\tfrac{1}{2})_{210}$ & $i\Phi[1,5,7,9] +\ldots$ \\
$C_{5}$ &$(1,2,2)_{120}$ &  $(0,0,0,0,\tfrac{3}{4},-\tfrac{1}{2},\tfrac{3}{4},+\tfrac{1}{2})_{120}$ & $-\Psi[1,2,3] +\ldots$ \\
$C_{6}$ & $(15,2,2)_{120}$ &  $(0,0,0,0,\tfrac{3}{4},-\tfrac{1}{2},\tfrac{3}{4},+\tfrac{1}{2})_{120}$ & $-\Psi[3,5,6] +\ldots$ \\
\hline \end{tabular}
\end{center}
\vskip 5mm
\paragraph{\bf Sector $(1,2,-\frac{3}{2})\oplus (1,2,+\frac{3}{2})$, $Q=\mp 2$ states:}
\be
 2 m_\Phi+6 \lambda  (a+\omega )
 \ee
\begin{center}
\begin{tabular}{|c|c|c|c|}
\hline 4 & Pati-Salam origin & ($C^{c}\!\!, T^{c}_{3},  T^{c}_{8}, B-L, C^{L}\!\!,T^{L}_{3}, C^{R}\!\!,  T^{R}_{3}$) &  phase convention\\
\hline
$R_{1}$ &$({10},2,2)_{210}$ &  $(0,0,0,-2,\tfrac{3}{4},-\tfrac{1}{2},\tfrac{3}{4},-\tfrac{1}{2})_{{210}}$ & $+i\Phi[1,5,7,9]+\ldots$ \\
\hline
$C_{1}$ &$(\overline{10},2,2)_{210}$ &  $(0,0,0,+2,\tfrac{3}{4},+\tfrac{1}{2},\tfrac{3}{4},+\tfrac{1}{2})_{210}$ & $-i\Phi[1,5,7,9] +\ldots$ \\
\hline \end{tabular}
\end{center}
and the results in the decoupling limit (i.e. $m_{\Phi}\to \infty$) are again the same as in \cite{Bajc:2004xe}, up to phase conventions.
\subsubsection{$C^{c}=0,  C^{L}=2$ - colourless $SU(2)_{L}$-triplets}
In the NMSGUT Higgs sector there are 3 colourless triplets belonging to this cathegory (accounting in total for 9 degrees of freedom) breaking into 2 subspaces with hypercharges $0$ and $\pm 1$. This sector is identical to the case of the minimal model.
\vskip 5mm
\paragraph{\bf Sector $(1,3,0)$ :}
\be
2 m_\Phi+2\lambda(2 a  -p)
\ee
\begin{center}
\begin{tabular}{|c|c|c|c|}
\hline 3 & Pati-Salam origin & ($C^{c}\!\!, T^{c}_{3},  T^{c}_{8}, B-L, C^{L}\!\!,T^{L}_{3}, C^{R}\!\!,  T^{R}_{3}$) &  phase convention\\
\hline
$S_{1}$ & $(15,3,1)_{210}$ & $(0,0,0,0,2,0,0,0)_{{210}}$ & $\Phi[1,2,5,6]+\ldots$ \\
\hline \end{tabular}
\end{center}
\vskip 5mm
\paragraph{\bf Sector $(1,3,-1)\oplus (1,3,+1)$, $Q=\mp 1$ states :}
\be
m_\Sigma+\eta(3 a  -p)
\ee
\begin{center}
\begin{tabular}{|c|c|c|c|}
\hline 6 & Pati-Salam origin & ($C^{c}\!\!, T^{c}_{3},  T^{c}_{8}, B-L, C^{L}\!\!,T^{L}_{3}, C^{R}\!\!,  T^{R}_{3}$) &  phase convention\\
\hline
$R_{1}$ & $({10},3,1)_{126}$ &  $(0,0,0,-2,2,0,0,0)_{{126}}$ & $-\Sigma[1,2,5,7,9] +\ldots$ \\
\hline
$C_{1}$ &  $(\overline{10},3,1)_{\overline{126}}$&  $(0,0,0,+2,2,0,0,0)_{\overline{126}}$ & $-\overline{\Sigma}[1,2,5,7,9] +\ldots$ \\
\hline \end{tabular}
\end{center}
As before, our results are the same like those given in \cite{Bajc:2004xe}.
\subsection{Mass matrices - colour triplets \& antitriplets}
The dimensionality of this part of the Higgs sector in the minimal model is 192. With addition of ${\bf 120}_{\rm H}$ there are extra 66 states in 11 extra triplet-antitriplet pairs yielding in total 258 degrees of freedom to be considered. For each of these triplets/antitriplets we shall consider only a representative corresponding to the $T^{c}_{3}=0$, $T^{c}_{8}=\pm \frac{1}{\sqrt{3}}$ states respectively.
\subsubsection{$C^{c}=\frac{4}{3},  C^{L}=0$ - colour triplets \& antitriplets, $SU(2)_{L}$-singlets}
There are four different types of multiplets transforming like colour triplets (and antitriplets) and $SU(2)_{L}$ singlets in the extended model accounting for 84 degrees of freedom in total -  according to the hypercharge they cluster into 4 sectors with $Y=\mp \frac{5}{3}$, $\mp\frac{2}{3}$, $\pm\frac{1}{3}$ and $\pm\frac{4}{3}$. Notice that the states of type $(\bar 3,1,-\frac{2}{3})\oplus (3,1,+\frac{2}{3})$ have the same quantum numbers like the gauge bosons from the $SO(10)/SU(3)_{c}\otimes SU(2)_{L}\otimes U(1)_{Y}$ coset (which must be made superheavy) and thus this part of the Higgs sector should again provide the relevant Goldstone bosons. For more detailed discussion of this point see Section \ref{goldstones}. 

Note also that the $(\bar 3,1,-\frac{4}{3})\oplus (3,1,+\frac{4}{3})$ and $(\bar 3,1,-\frac{1}{3})\oplus (3,1,+\frac{1}{3})$ sectors are in general proton-decay dangerous (inducing in SUSY the $d=5$ operators $\hat U^{c}\hat U^{c}\hat D^{c}\hat E^{c}$ and $\hat Q\hat Q\hat Q\hat L$ respectively) and thus all the eigenvalues should live at least at the GUT scale in potentially realistic settings (with SUSY at around 1 TeV).
\vskip 5mm
\paragraph{\bf Sector $(\bar 3,1,-\frac{5}{3})\oplus (3,1,+\frac{5}{3})$, $T^{c}_{3}=0$, $T^{c}_{8}=\pm \frac{1}{\sqrt{3}}$ states:}
\be
 2 (m_\Phi+\lambda  (a+p+4 \omega ))
\ee
\begin{center}
\begin{tabular}{|c|c|c|c|}
\hline 6 & Pati-Salam origin & ($C^{c}\!\!, T^{c}_{3},  T^{c}_{8}, B-L, C^{L}\!\!,T^{L}_{3}, C^{R}\!\!,  T^{R}_{3}$) &  phase convention\\
\hline
$R_{1}$ & $(15,1,3)_{210}$ &  $(\frac{4}{3},0,+\frac{1}{\sqrt{3}},-\frac{4}{3},0,0,2,-1)_{{210}}$ & $+i\Phi[1,3,5,7] +\ldots$ \\
\hline
$C_{1}$ & $(15,1,3)_{210}$ &  $(\frac{4}{3},0,-\frac{1}{\sqrt{3}},+\frac{4}{3},0,0,2,+1)_{\overline{210}}$ & $-i\Phi[1,3,5,7] +\ldots$ \\
\hline \end{tabular}
\end{center}
\vskip 5mm
\paragraph{\bf Sector $(\bar 3,1,-\frac{2}{3})\oplus (3,1,+\frac{2}{3})$, $T^{c}_{3}=0$, $T^{c}_{8}=\pm \frac{1}{\sqrt{3}}$ states:}
\be
\left(
\begin{array}{cccc}
 m_\Sigma+\eta  (a+p-2 \omega ) & \sqrt{2} \eta  \overline{\sigma} & -2 \eta  \overline{\sigma}
   & -2 \sqrt{2} \beta  (a-2 \omega ) \\
 \sqrt{2} \eta  \sigma  & 2 (m_\Phi+a \lambda ) & 2 \sqrt{2} \lambda  \omega  & 2 \beta  \sigma  \\
 -2 \eta  \sigma  & 2 \sqrt{2} \lambda  \omega  & 2 (m_\Phi+(a+p) \lambda ) & -2 \sqrt{2} \beta 
   \sigma  \\
 -2 \sqrt{2} \bar\beta (a-2 \omega ) & 2 \bar\beta \overline{\sigma} & -2 \sqrt{2}
   \bar\beta \overline{\sigma} & 2 (m_\Psi+\rho  (p-2 \omega ))
\end{array}
\right)
\ee
\begin{center}
\begin{tabular}{|c|c|c|c|}
\hline 24 & Pati-Salam origin & ($C^{c}\!\!, T^{c}_{3},  T^{c}_{8}, B-L, C^{L}\!\!,T^{L}_{3}, C^{R}\!\!,  T^{R}_{3}$) &  phase convention\\
\hline
$R_{1}$ & $(\overline{10},1,3)_{126}$&  $(\frac{4}{3},0,+\frac{1}{\sqrt{3}},+\frac{2}{3},0,0,2,-1)_{{126}}$ & $\Sigma[1,3,5,6,9] +\ldots$ \\
$R_{2}$ & $(15,1,1)_{210}$ &  $(\frac{4}{3},0,+\frac{1}{\sqrt{3}},-\frac{4}{3},0,0,0,0)_{{210}}$ & $-\Phi[5,7,9,10] +\ldots$ \\
$R_{3}$ & $(15,1,3)_{210}$ &  $(\frac{4}{3},0,+\frac{1}{\sqrt{3}},-\frac{4}{3},0,0,2,0)_{{210}}$ & $-\Phi[1,2,5,7] +\ldots$ \\
$R_{4}$ & $(6,1,3)_{120}$ &  $(\frac{4}{3},0,+\frac{1}{\sqrt{3}},+\frac{2}{3},0,0,2,-1)_{{120}}$ & $-i\Psi[1,3,9] +\ldots$ \\
\hline
$C_{1}$ & $({10},1,3)_{\overline{126}}$&  $(\frac{4}{3},0,-\frac{1}{\sqrt{3}},-\frac{2}{3},0,0,2,+1)_{\overline{126}}$ & $\overline{\Sigma}[1,3,5,6,9]  +\ldots$ \\
$C_{2}$ & $(15,1,1)_{210}$ &  $(\frac{4}{3},0,-\frac{1}{\sqrt{3}},+\frac{4}{3},0,0,0,0)_{{210}}$ & $-\Phi[5,7,9,10] +\ldots$ \\
$C_{3}$ & $(15,1,3)_{210}$ &  $(\frac{4}{3},0,-\frac{1}{\sqrt{3}},+\frac{4}{3},0,0,2,0)_{{210}}$ & $-\Phi[1,2,5,7]  +\ldots$ \\
$C_{4}$ & $(6,1,3)_{120}$ &  $(\frac{4}{3},0,-\frac{1}{\sqrt{3}},-\frac{2}{3},0,0,2,+1)_{{120}}$ & $+i\Psi[1,3,9] +\ldots$ \\
\hline \end{tabular}
\end{center}
\vskip 5mm
\paragraph{\bf Sector $(\bar 3,1,+\frac{1}{3})\oplus (3,1,-\frac{1}{3})$, $T^{c}_{3}=0$, $T^{c}_{8}=\pm \frac{1}{\sqrt{3}}$ states: :}
\be
\left(
\begin{array}{ccccccc}
 2m_H& \frac{1}{\sqrt{2}} (a+p) \bar\alpha& \frac{1}{\sqrt{2}}(p-a) \alpha  & 2\bar\alpha \omega  & -\bar\alpha \overline{\sigma}  & \sqrt{2} a \gamma  & \sqrt{2} \gamma  \omega 
   \\
 \frac{1}{\sqrt{2}}(a+p) \alpha  & m_\Sigma & 0 & 2 \sqrt{2} \eta  \omega  & -\sqrt{2} \eta 
   \overline{\sigma} & 2 a \beta  & 2 \beta  \omega  \\
 \frac{1}{\sqrt{2}}(p-a) \bar\alpha & 0 & m_\Sigma & 0 & 0 & -2 a \bar\beta & 2
   \bar\beta \omega  \\
 2 \alpha  \omega  & 2 \sqrt{2} \eta  \omega  & 0 & m_\Sigma+(a+p) \eta  & 2 \eta  \overline{\sigma} & 2 \sqrt{2} \beta  \omega  & 2 \sqrt{2} a \beta  \\
 -\alpha  \sigma  & -\sqrt{2} \eta  \sigma  & 0 & 2 \eta  \sigma  & 2 (m_\Phi+\lambda  (a+p-4
   \omega )) & 2 \sqrt{2} \beta  \sigma  & -2 \sqrt{2} \beta  \sigma  \\
 \sqrt{2} a \gamma  & 2 a \bar\beta & -2 a \beta  & 2 \sqrt{2} \bar\beta \omega  & 2
   \sqrt{2} \bar\beta \overline{\sigma} & 2 (m_\Psi+a \rho ) & 4 \rho  \omega  \\
 \sqrt{2} \gamma  \omega  & 2 \bar\beta \omega  & 2 \beta  \omega  & 2 \sqrt{2} a \bar\beta & -2 \sqrt{2} \bar\beta \overline{\sigma} & 4 \rho  \omega  & 2 (m_\Psi+p \rho
   )
\end{array}
\right)
\ee
\begin{center}
\begin{tabular}{|c|c|c|c|}
\hline 42 & Pati-Salam origin & ($C^{c}\!\!, T^{c}_{3},  T^{c}_{8}, B-L, C^{L}\!\!,T^{L}_{3}, C^{R}\!\!,  T^{R}_{3}$) &  phase convention\\
\hline
$R_{1}$ & $(6,1,1)_{10}$ &  $(\frac{4}{3},0,+\frac{1}{\sqrt{3}},+\frac{2}{3},0,0,0,0)_{{10}}$ & $ +iH[9]+\ldots$ \\
$R_{2}$ & $(6,1,1)_{126}$&  $(\frac{4}{3},0,+\frac{1}{\sqrt{3}},+\frac{2}{3},0,0,0,0)_{{126}}$ & $ +i\Sigma[1,2,3,4,9]+\ldots$ \\
$R_{3}$ & $(6,1,1)_{\overline{126}}$&  $(\frac{4}{3},0,+\frac{1}{\sqrt{3}},+\frac{2}{3},0,0,0,0)_{\overline{126}}$ & $+i\overline{\Sigma}[1,2,3,4,9] +\ldots$ \\
$R_{4}$ & $(\overline{10},1,3)_{{126}}$&  $(\frac{4}{3},0,+\frac{1}{\sqrt{3}},+\frac{2}{3},0,0,2,0)_{{126}}$ & $i\Sigma[1,2,5,6,9] +\ldots$ \\
$R_{5}$ &$(15,1,3)_{210}$ &  $(\frac{4}{3},0,+\frac{1}{\sqrt{3}},-\frac{4}{3},0,0,2,+1)_{{210}}$ & $-i \Phi[1,3,5,7] +\ldots$ \\
$R_{6}$ & $(\overline{10},1,1)_{{120}}$&  $(\frac{4}{3},0,+\frac{1}{\sqrt{3}},+\frac{2}{3},0,0,0,0)_{{120}}$ & $-\Psi[5,6,9] +\ldots$ \\
$R_{7}$ & $(6,1,3)_{120}$&  $(\frac{4}{3},0,+\frac{1}{\sqrt{3}},+\frac{2}{3},0,0,2,0)_{{120}}$ & $-\Psi[1,2,9] +\ldots$ \\
\hline
$C_{1}$ & $(6,1,1)_{10}$ &  $(\frac{4}{3},0,-\frac{1}{\sqrt{3}},-\frac{2}{3},0,0,0,0)_{{10}}$ & $-iH[9] +\ldots$ \\
$C_{2}$ & $(6,1,1)_{\overline{126}}$&  $(\frac{4}{3},0,-\frac{1}{\sqrt{3}},-\frac{2}{3},0,0,0,0)_{\overline{126}}$ & $ -i\overline{\Sigma}[1,2,3,4,9] +\ldots$ \\
$C_{3}$ & $(6,1,1)_{126}$&  $(\frac{4}{3},0,-\frac{1}{\sqrt{3}},-\frac{2}{3},0,0,0,0)_{{126}}$ & $-i\Sigma[1,2,3,4,9] +\ldots$ \\
$C_{4}$ &  $({10},1,3)_{\overline{126}}$&  $(\frac{4}{3},0,-\frac{1}{\sqrt{3}},-\frac{2}{3},0,0,2,0)_{{126}}$ & $-i\overline{\Sigma}[1,2,5,6,9] +\ldots$ \\
$C_{5}$ & $(15,1,3)_{210}$&  $(\frac{4}{3},0,-\frac{1}{\sqrt{3}},+\frac{4}{3},0,0,2,-1)_{{210}}$ & $+i \Phi[1,3,5,7] +\ldots$ \\
$C_{6}$ & $({10},1,1)_{{120}}$&  $(\frac{4}{3},0,-\frac{1}{\sqrt{3}},-\frac{2}{3},0,0,0,0)_{{120}}$ & $-\Psi[5,6,9] +\ldots$ \\
$C_{7}$ & $(6,1,3)_{120}$&  $(\frac{4}{3},0,-\frac{1}{\sqrt{3}},-\frac{2}{3},0,0,2,0)_{{120}}$ & $-\Psi[1,2,9] +\ldots$ \\
\hline \end{tabular}
\end{center}
\vskip 5mm
\paragraph{\bf Sector $(\bar 3,1,+\frac{4}{3})\oplus (3,1,-\frac{4}{3})$, $T^{c}_{3}=0$, $T^{c}_{8}=\pm \frac{1}{\sqrt{3}}$ states: :}
\be
\left(
\begin{array}{cc}
 m_\Sigma+\eta  (a+p+2 \omega ) & 2 \sqrt{2} \beta  (a+2 \omega ) \\
 2 \sqrt{2} \bar\beta (a+2 \omega ) & 2 (m_\Psi+\rho  (p+2 \omega ))
\end{array}
\right)
\ee
\begin{center}
\begin{tabular}{|c|c|c|c|}
\hline 12 & Pati-Salam origin & ($C^{c}\!\!, T^{c}_{3},  T^{c}_{8}, B-L, C^{L}\!\!,T^{L}_{3}, C^{R}\!\!,  T^{R}_{3}$) &  phase convention\\
\hline
$R_{1}$ & $(\overline{10},1,3)_{126}$&  $(\frac{4}{3},0,+\frac{1}{\sqrt{3}},+\frac{2}{3},0,0,2,+1)_{{126}}$ & $  \Sigma[1,3,5,6,9]+\ldots$ \\
$R_{2}$ & $(6,1,3)_{120}$&  $(\frac{4}{3},0,+\frac{1}{\sqrt{3}},+\frac{2}{3},0,0,2,+1)_{{120}}$ & $ +i\Psi[1,3,9]+\ldots$ \\
\hline
$C_{1}$ & $({10},1,3)_{\overline{126}}$&  $(\frac{4}{3},0,-\frac{1}{\sqrt{3}},-\frac{2}{3},0,0,2,-1)_{\overline{126}}$ & $ \overline{\Sigma}[1,3,5,6,9] +\ldots$ \\
$C_{2}$ &  $(6,1,3)_{120}$ &  $(\frac{4}{3},0,-\frac{1}{\sqrt{3}},-\frac{2}{3},0,0,2,-1)_{{120}}$ & $ -i\Psi[1,3,9]+\ldots$ \\
\hline \end{tabular}
\end{center}
\vskip 5mm
As before, it can be checked our results correspond to those of \cite{Bajc:2004xe} wherever appropriate.
\subsubsection{$C^{c}=\frac{4}{3},  C^{L}=\frac{3}{4}$ - colour triplets \& antitriplets, $SU(2)_{L}$-doublets}
This sector consists of the total of 120 states clustered into 10 triplet-antitriplet pairs spanning $SU(2)_{L}$ doublets with hypercharges $Y=\mp \frac{7}{6}$, $\mp \frac{1}{6}$ and $\pm\frac{5}{6}$. The piece of particular interest is the one corresponding to the  $(\bar 3,2,-\frac{1}{6})\oplus (3,2,+\frac{1}{6})$  quantum numbers as it should, as before, provide the Goldstone bosons for the relevant gauge sector to become massive; for details see Section \ref{goldstones}. Furthermore, the $(\bar 3,2,+\frac{5}{6})\oplus (3,2,-\frac{5}{6})$ fields can mix with the relevant gauginos and participate at the proton decay. 
\vskip 5mm
\paragraph{\bf Sector $(\bar 3,2,-\frac{7}{6})\oplus (3,2,+\frac{7}{6})$, $T^{c}_{3}=0$, $T^{c}_{8}=\pm \frac{1}{\sqrt{3}}$ states: :}
\be
\left(
\begin{array}{ccc}
 m_\Sigma+\eta  (a+\omega ) & 0 & \sqrt{2} \beta  (2 a-p-\omega ) \\
 0 & m_\Sigma+\eta  (a+3 \omega ) & -\sqrt{2} \bar\beta (2 a+p+3 \omega ) \\
 \sqrt{2} \bar\beta (2 a-p-\omega ) & -\sqrt{2} \beta  (2 a+p+3 \omega ) & 2 (m_\Psi+\rho 
   (a+2 \omega ))
\end{array}
\right)
\ee
\begin{center}
\begin{tabular}{|c|c|c|c|}
\hline 36 & Pati-Salam origin & ($C^{c}\!\!, T^{c}_{3},  T^{c}_{8}, B-L, C^{L}\!\!,T^{L}_{3}, C^{R}\!\!,  T^{R}_{3}$) &  phase convention\\
\hline
$R_{1}$ & $(15,2,2)_{126}$&  $(\frac{4}{3},0,+\frac{1}{\sqrt{3}},-\frac{4}{3},\frac{3}{4},+\frac{1}{2},\frac{3}{4},-\frac{1}{2})_{{126}}$ & $+i\Sigma[1,2,3,5,7]+\ldots$ \\
$R_{2}$ & $(15,2,2)_{\overline{126}}$&  $(\frac{4}{3},0,+\frac{1}{\sqrt{3}},-\frac{4}{3},\frac{3}{4},+\frac{1}{2},\frac{3}{4},-\frac{1}{2})_{\overline{126}}$ & $ +i\overline{\Sigma}[1,2,3,5,7]+\ldots$ \\
$R_{3}$ &  $(15,2,2)_{120}$&  $(\frac{4}{3},0,+\frac{1}{\sqrt{3}},-\frac{4}{3},\frac{3}{4},+\frac{1}{2},\frac{3}{4},-\frac{1}{2})_{{120}}$ & $ -\Psi[3,5,7]+\ldots$ \\
\hline
$C_{1}$ &  $(15,2,2)_{\overline{126}}$&  $(\frac{4}{3},0,-\frac{1}{\sqrt{3}},+\frac{4}{3},\frac{3}{4},-\frac{1}{2},\frac{3}{4},+\frac{1}{2})_{\overline{126}}$ & $ -i\overline{\Sigma}[1,2,3,5,7]+\ldots$ \\
$C_{2}$ & $(15,2,2)_{126}$&  $(\frac{4}{3},0,-\frac{1}{\sqrt{3}},+\frac{4}{3},\frac{3}{4},-\frac{1}{2},\frac{3}{4},+\frac{1}{2})_{{126}}$ & $-i\Sigma[1,2,3,5,7] +\ldots$ \\
$C_{3}$ &  $(15,2,2)_{120}$ &  $(\frac{4}{3},0,-\frac{1}{\sqrt{3}},+\frac{4}{3},\frac{3}{4},-\frac{1}{2},\frac{3}{4},+\frac{1}{2})_{{120}}$ & $-\Psi[3,5,7] +\ldots$ \\
\hline \end{tabular}
\end{center}
\vskip 5mm
\paragraph{\bf Sector $(\bar 3,2,-\frac{1}{6})\oplus (3,2,+\frac{1}{6})$, $T^{c}_{3}=0$, $T^{c}_{8}=\pm \frac{1}{\sqrt{3}}$ states: :}
\be
\left(
\begin{array}{ccccc}
 m_\Sigma+\eta  (a-\omega ) & 0 & 0 & 0 & -\sqrt{2} \beta  (2 a-p+\omega ) \\
 0 & m_\Sigma+\eta  (a-3 \omega ) & -\sqrt{2} \eta  \sigma  & 2 \eta  \sigma  & -\sqrt{2}
   \bar\beta (2 a+p-3 \omega ) \\
 0 & -\sqrt{2} \eta  \overline{\sigma} & 2 (m_\Phi-\lambda  \omega ) & 2 \sqrt{2} \lambda  \omega
    & -2 \bar\beta \overline{\sigma} \\
 0 & 2 \eta  \overline{\sigma} & 2 \sqrt{2} \lambda  \omega  & 2 (m_\Phi+\lambda  (a-\omega )) &
   2 \sqrt{2} \bar\beta \overline{\sigma} \\
 -\sqrt{2} \bar\beta (2 a-p+\omega ) & -\sqrt{2} \beta  (2 a+p-3 \omega ) & -2 \beta  \sigma  & 2
   \sqrt{2} \beta  \sigma  & 2 (m_\Psi+\rho  (a-2 \omega ))
\end{array}
\right)
\ee
\begin{center}
\begin{tabular}{|c|c|c|c|}
\hline 60 & Pati-Salam origin & ($C^{c}\!\!, T^{c}_{3},  T^{c}_{8}, B-L, C^{L}\!\!,T^{L}_{3}, C^{R}\!\!,  T^{R}_{3}$) &  phase convention\\
\hline
$R_{1}$ &  $(15,2,2)_{126}$&  $(\frac{4}{3},0,+\frac{1}{\sqrt{3}},-\frac{4}{3},\frac{3}{4},+\frac{1}{2},\frac{3}{4},+\frac{1}{2})_{{126}}$ & $-\Sigma[1,3,4,5,7] +\ldots$ \\
$R_{2}$ &  $(15,2,2)_{\overline{126}}$ &  $(\frac{4}{3},0,+\frac{1}{\sqrt{3}},-\frac{4}{3},\frac{3}{4},+\frac{1}{2},\frac{3}{4},+\frac{1}{2})_{\overline{126}}$ & $ \overline{\Sigma}[1,3,4,5,7]+\ldots$ \\
$R_{3}$ & $(6,2,2)_{210}$&  $(\frac{4}{3},0,+\frac{1}{\sqrt{3}},+\frac{2}{3},\frac{3}{4},+\frac{1}{2},\frac{3}{4},-\frac{1}{2})_{{210}}$ & $\Phi[1,2,3,9] +\ldots$ \\
$R_{4}$ &  $(\overline{10},2,2)_{210}$&  $(\frac{4}{3},0,+\frac{1}{\sqrt{3}},+\frac{2}{3},\frac{3}{4},+\frac{1}{2},\frac{3}{4},-\frac{1}{2})_{{210}}$ & $ \Phi[3,5,6,9]+\ldots$ \\
$R_{5}$ &  $(15,2,2)_{120}$&  $(\frac{4}{3},0,+\frac{1}{\sqrt{3}},-\frac{4}{3},\frac{3}{4},+\frac{1}{2},\frac{3}{4},+\frac{1}{2})_{{120}}$ & $-i \Psi [1,5,7] +\ldots$ \\
\hline
$C_{1}$ &  $(15,2,2)_{\overline{126}}$&  $(\frac{4}{3},0,-\frac{1}{\sqrt{3}},+\frac{4}{3},\frac{3}{4},-\frac{1}{2},\frac{3}{4},-\frac{1}{2})_{\overline{126}}$ & $-\overline{\Sigma}[1,3,4,5,7] +\ldots$ \\
$C_{2}$ & $(15,2,2)_{126}$ &  $(\frac{4}{3},0,-\frac{1}{\sqrt{3}},+\frac{4}{3},\frac{3}{4},-\frac{1}{2},\frac{3}{4},-\frac{1}{2})_{{126}}$ & $\Sigma[1,3,4,5,7]+\ldots$ \\
$C_{3}$ &  $(6,2,2)_{210}$&  $(\frac{4}{3},0,-\frac{1}{\sqrt{3}},-\frac{2}{3},\frac{3}{4},-\frac{1}{2},\frac{3}{4},+\frac{1}{2})_{{210}}$ & $ \Phi[1,2,3,9]+\ldots$ \\
$C_{4}$ &  $({10},2,2)_{210}$&  $(\frac{4}{3},0,-\frac{1}{\sqrt{3}},-\frac{2}{3},\frac{3}{4},-\frac{1}{2},\frac{3}{4},+\frac{1}{2})_{{210}}$ & $ \Phi[3,5,6,9]+\ldots$ \\
$C_{5}$ &  $(15,2,2)_{120}$&  $(\frac{4}{3},0,-\frac{1}{\sqrt{3}},+\frac{4}{3},\frac{3}{4},-\frac{1}{2},\frac{3}{4},-\frac{1}{2})_{{120}}$ & $ +i \Psi [1,5,7]+\ldots$ \\
\hline \end{tabular}
\end{center}
\vskip 5mm
\paragraph{\bf Sector $(\bar 3,2,+\frac{5}{6})\oplus (3,2,-\frac{5}{6})$, $T^{c}_{3}=0$, $T^{c}_{8}=\pm \frac{1}{\sqrt{3}}$ states: :}
\be
\left(
\begin{array}{cc}
 2 (m_\Phi+\lambda  \omega ) & 2 \sqrt{2} \lambda  \omega  \\
 2 \sqrt{2} \lambda  \omega  & 2 (m_\Phi+\lambda  (a+\omega ))
\end{array}
\right)
\ee
\begin{center}
\begin{tabular}{|c|c|c|c|}
\hline 24 & Pati-Salam origin & ($C^{c}\!\!, T^{c}_{3},  T^{c}_{8}, B-L, C^{L}\!\!,T^{L}_{3}, C^{R}\!\!,  T^{R}_{3}$) &  phase convention\\
\hline
$R_{1}$ &  $(6,2,2)_{210}$&  $(\frac{4}{3},0,+\frac{1}{\sqrt{3}},+\frac{2}{3},\frac{3}{4},+\frac{1}{2},\frac{3}{4},+\frac{1}{2})_{{210}}$ & $-\Phi[1,3,4,9] +\ldots$ \\
$R_{2}$ & $(\overline{10},2,2)_{210}$ &  $(\frac{4}{3},0,+\frac{1}{\sqrt{3}},+\frac{2}{3},\frac{3}{4},+\frac{1}{2},\frac{3}{4},+\frac{1}{2})_{{210}}$ & $-\Phi[1,5,6,9]  +\ldots$ \\
\hline
$C_{1}$ &  $(6,2,2)_{210}$&  $(\frac{4}{3},0,-\frac{1}{\sqrt{3}},-\frac{2}{3},\frac{3}{4},-\frac{1}{2},\frac{3}{4},-\frac{1}{2})_{{210}}$ & $ -\Phi[1,3,4,9]+\ldots$ \\
$C_{2}$ &$({10},2,2)_{210}$ &  $(\frac{4}{3},0,-\frac{1}{\sqrt{3}},-\frac{2}{3},\frac{3}{4},-\frac{1}{2},\frac{3}{4},-\frac{1}{2})_{{210}}$ & $-\Phi[1,5,6,9]  +\ldots$ \\
\hline \end{tabular}
\end{center}
\vskip 5mm
\subsubsection{$C^{c}=\frac{4}{3},  C^{L}=2$ - colour triplets \& antitriplets, $SU(2)_{L}$-triplets}
The coloured triplet-antitriplet pairs that transform like triplets under $SU(2)_{L}$ account in total for 54 states with hypercharges $Y=\mp\frac{2}{3}$ and $\pm \frac{1}{3}$. The only difference between the minimal model and the extended scenario is one extra set of states in the $Y\pm \frac{1}{3}$ sector though.
\vskip 5mm 
\paragraph{\bf Sector $(\bar 3,3,-\frac{2}{3})\oplus (3,3,+\frac{2}{3})$, $T^{c}_{3}=0$, $T^{c}_{8}=\pm \frac{1}{\sqrt{3}}$ states: :}
\be
2 (m_\Phi+(a-p) \lambda )
\ee
\begin{center}
\begin{tabular}{|c|c|c|c|}
\hline 18 & Pati-Salam origin & ($C^{c}\!\!, T^{c}_{3},  T^{c}_{8}, B-L, C^{L}\!\!,T^{L}_{3}, C^{R}\!\!,  T^{R}_{3}$) &  phase convention\\
\hline
$R_{1}$ & $(15,3,1)_{210}$ &  $(\frac{4}{3},0,+\frac{1}{\sqrt{3}},-\frac{4}{3},2,0,0,0)_{{210}}$ & $-\Phi[1,2,5,7] +\ldots$ \\
\hline
$C_{1}$ &$(15,3,1)_{210}$ &  $(\frac{4}{3},0,-\frac{1}{\sqrt{3}},+\frac{4}{3},2,0,0,0)_{{210}}$ & $-\Phi[1,2,5,7] +\ldots$ \\
\hline \end{tabular}
\end{center}
\vskip 5mm
\paragraph{\bf Sector $(\bar 3,3,+\frac{1}{3})\oplus (3,3,-\frac{1}{3})$, $T^{c}_{3}=0$, $T^{c}_{8}=\pm \frac{1}{\sqrt{3}}$ states: :}
\be
\left(
\begin{array}{cc}
 m_\Sigma+(a-p) \eta  & 2 \sqrt{2} a \bar\beta \\
 2 \sqrt{2} a \beta  & 2 (m_\Psi-p \rho )
\end{array}
\right)
\ee
\begin{center}
\begin{tabular}{|c|c|c|c|}
\hline 36 & Pati-Salam origin & ($C^{c}\!\!, T^{c}_{3},  T^{c}_{8}, B-L, C^{L}\!\!,T^{L}_{3}, C^{R}\!\!,  T^{R}_{3}$) &  phase convention\\
\hline
$R_{1}$ & $(\overline{10},3,1)_{\overline{126}}$&  $(\frac{4}{3},0,+\frac{1}{\sqrt{3}},+\frac{2}{3},2,0,0,0)_{\overline{126}}$ & $\overline{\Sigma}[1,2,5,6,9] +\ldots$ \\
$R_{2}$ & $(6,3,1)_{120}$ &  $(\frac{4}{3},0,+\frac{1}{\sqrt{3}},+\frac{2}{3},2,0,0,0)_{{120}}$ & $+i\Psi[1,2,9] +\ldots$ \\
\hline
$C_{1}$ &  $({10},3,1)_{{126}}$&  $(\frac{4}{3},0,-\frac{1}{\sqrt{3}},-\frac{2}{3},2,0,0,0)_{{126}}$ & $\Sigma[1,2,5,6,9]  +\ldots$ \\
$C_{2}$ & $(6,3,1)_{120}$ &  $(\frac{4}{3},0,-\frac{1}{\sqrt{3}},-\frac{2}{3},2,0,0,0)_{{120}}$ & $-i\Psi[1,2,9] +\ldots$ \\
\hline \end{tabular}
\end{center}
\vskip 5mm
As before, in the $m_{\Psi}\to \infty$ decoupling limit we do reconstruct the result of the minimal model study \cite{Bajc:2004xe}. 
\subsection{Mass matrices - colour octets}
There is in total 152 states  in 19 different colour octets belonging into this cathegory, to be compared with 15 such octets in the minimal model. The new states from ${\bf 120}_{\rm H}$ enter only the $(8,2,-\frac{1}{2})\oplus (8,2,+\frac{1}{2})$ sector while the rest is identical to the MSGUT situation. 
\subsubsection{$C^{c}=3,  C^{L}=0$ - colour octets, $SU(2)_{L}$-singlets}
This subset is entirely identical to the corresponding one in the minimal model so there is no need to further comment on it. 
\paragraph{\bf Sector $(8,1,-1)\oplus (8,1,+1)$, $T^{c}_{3}=\mp 1$, $T^{c}_{8}=0$ states:}
\be
2 (m_\Phi+(p-a) \lambda )
\ee
\begin{center}
\begin{tabular}{|c|c|c|c|}
\hline 16 & Pati-Salam origin & ($C^{c}\!\!, T^{c}_{3},  T^{c}_{8}, B-L, C^{L}\!\!,T^{L}_{3}, C^{R}\!\!,  T^{R}_{3}$) &  phase convention\\
\hline
$R_{1}$ & $(15,1,3)_{210}$&  $(3,-1,0,0,0,0,2,-1)_{{210}}$ & $\Phi[1, 3, 5, 7] +\ldots$ \\
\hline
$C_{1}$ & $(15,1,3)_{210}$ &  $(3,+1,0,0,0,0,2,+1)_{{210}}$ & $\Phi[1, 3, 5, 7] +\ldots$ \\
\hline \end{tabular}
\end{center}
\vskip 5mm
\paragraph{\bf Sector $(8,1,0)$, $T^{c}_{3}=\mp 1$, $T^{c}_{8}=0$ states:}
\be
\left(
\begin{array}{cc}
 2 (m_\Phi-a \lambda ) & 2 \sqrt{2} \lambda  \omega  \\
 2 \sqrt{2} \lambda  \omega  & 2 (m_\Phi+(p-a) \lambda )
\end{array}
\right)
\ee
\begin{center}
\begin{tabular}{|c|c|c|c|}
\hline 16 & Pati-Salam origin & ($C^{c}\!\!, T^{c}_{3},  T^{c}_{8}, B-L, C^{L}\!\!,T^{L}_{3}, C^{R}\!\!,  T^{R}_{3}$) &  phase convention\\
\hline
$R_{1}$ & $(15,1,1)_{210}$&  $(3,-1,0,0,0,0,0,0)_{{210}}$ & $+i\Phi[5, 7, 9, 10] +\ldots$ \\
$R_{2}$ & $(15,1,3)_{210}$&  $(3,-1,0,0,0,0,2,0)_{{210}}$ & $+i\Phi[1,2,5, 7] +\ldots$ \\
\hline
$C_{1}$ & $(15,1,1)_{210}$&  $(3,+1,0,0,0,0,0,0)_{{210}}$ & $-i\Phi[5, 7, 9, 10] +\ldots$ \\
$C_{2}$ & $(15,1,3)_{210}$&  $(3,+1,0,0,0,0,2,0)_{{210}}$ & $-i\Phi[1,2,5, 7]  +\ldots$ \\
\hline \end{tabular}
\end{center}
\vskip 5mm
\subsubsection{$C^{c}=3,  C^{L}=\frac{3}{4}$ - colour octets, $SU(2)_{L}$-doublets}
As anticipated, the only change due to the new multiplet in the game in the coloured octet subspace propagates into an extra row/column in the following sector:
\vskip 5mm 
\paragraph{\bf Sector $(8,2,-\frac{1}{2})\oplus (8,2,+\frac{1}{2})$, $T^{c}_{3}=\mp 1$, $T^{c}_{8}=0$ states:}
\be
\left(
\begin{array}{lll}
 m_\Sigma+\eta  (\omega -a) & 0 & \sqrt{2} \beta  (p+\omega ) \\
 0 & m_\Sigma-\eta  (a+\omega ) & \sqrt{2} \bar\beta (\omega -p) \\
 \sqrt{2} \bar\beta (p+\omega ) & \sqrt{2} \beta  (\omega -p) & 2 (m_\Psi-a \rho )
\end{array}
\right)
\ee
\begin{center}
\begin{tabular}{|c|c|c|c|}
\hline 96 & Pati-Salam origin & ($C^{c}\!\!, T^{c}_{3},  T^{c}_{8}, B-L, C^{L}\!\!,T^{L}_{3}, C^{R}\!\!,  T^{R}_{3}$) &  phase convention\\
\hline
$R_{1}$ &  $(15,2,2)_{126}$&  $(3,-1,0,0,\frac{3}{4},-\frac{1}{2},\frac{3}{4},-\frac{1}{2})_{{126}}$ & $+i\Sigma[1, 3, 4, 5, 7] +\ldots$ \\
$R_{2}$ & $(15,2,2)_{\overline{126}}$&  $(3,-1,0,0,\frac{3}{4},-\frac{1}{2},\frac{3}{4},-\frac{1}{2})_{\overline{126}}$ & $-i\overline{\Sigma}[1, 3, 4, 5, 7] +\ldots$ \\
$R_{3}$ & $(15,2,2)_{120}$&  $(3,-1,0,0,\frac{3}{4},-\frac{1}{2},\frac{3}{4},-\frac{1}{2})_{{120}}$ & $\Psi [1,5,7] +\ldots$ \\
\hline
$C_{1}$ & $(15,2,2)_{\overline{126}}$&  $(3,+1,0,0,\frac{3}{4},+\frac{1}{2},\frac{3}{4},+\frac{1}{2})_{\overline{126}}$ & $-i\overline{\Sigma}[1, 3, 4, 5, 7] +\ldots$ \\
$C_{2}$ & $(15,2,2)_{126}$&  $(3,+1,0,0,\frac{3}{4},+\frac{1}{2},\frac{3}{4},+\frac{1}{2})_{{126}}$ & $+i\Sigma[1, 3, 4, 5, 7]  +\ldots$ \\
$C_{3}$ & $(15,2,2)_{120}$&  $(3,+1,0,0,\frac{3}{4},+\frac{1}{2},\frac{3}{4},+\frac{1}{2})_{{120}}$ & $\Psi [1,5,7] +\ldots$ \\
\hline \end{tabular}
\end{center}
\vskip 5mm
\subsubsection{$C^{c}=3,  C^{L}=2$ - colour octets, $SU(2)_{L}$-triplets}
This sector is again identical to the corresponding minimal model one, so there is no need for extra comments.
\vskip 5mm
\paragraph{\bf Sector $(8,3,0)$, $T^{c}_{3}=\mp 1$, $T^{c}_{8}=0$ states:}
\be
2 (m_\Phi-(a+p) \lambda )
\ee
\begin{center}
\begin{tabular}{|c|c|c|c|}
\hline 24 & Pati-Salam origin & ($C^{c}\!\!, T^{c}_{3},  T^{c}_{8}, B-L, C^{L}\!\!,T^{L}_{3}, C^{R}\!\!,  T^{R}_{3}$) &  phase convention\\
\hline
$R_{1}$ &  $(15,3,1)_{210}$&  $(3,-1,0,0,2,0,0,0)_{{210}}$ & $+i \Phi[1, 2, 5, 7] +\ldots$ \\
\hline
$C_{1}$ &  $(15,3,1)_{210}$&  $(3,+1,0,0,2,0,0,0)_{{210}}$ & $-i \Phi[1, 2, 5, 7] +\ldots$ \\
\hline \end{tabular}
\end{center}
\vskip 5mm
\subsection{Mass matrices - colour sextets \& antisextets}
There are 132 degrees of freedom corresponding to the colour sextet-antisextet pairs in the next to minimal SUSY $SO(10)$ (to be compared to 120 in the MSGUT \cite{Bajc:2004xe}). Since all sextets descend from the Pati-Salam decuplets decomposing under $SU(3)\otimes U(1)_{B-L}$ like ${\bf 10}=(6,-\frac{2}{3})\oplus (3,+\frac{2}{3})\oplus (1,+2)$ and the only such decuplets within the extra ${\bf 120}_{\rm H}$ are $SU(2)_{L}\otimes SU(2)_{R}$ singlets, the minimal model situation can be  altered solely in the $(6,1,-\frac{1}{3})\oplus (\overline{6},1,+\frac{2}{3})$ sector.
\subsubsection{$C^{c}=\frac{10}{3},  C^{L}=0$ - colour sextets \& antisextets, $SU(2)_{L}$-singlets}
The only difference with respect to the MSGUT coloured sextets is that the $(6,1,-\frac{1}{3})\oplus (\overline{6},1,+\frac{2}{3})$ states are spanned over more than just a single $SO(10)$ multiplet which promotes the relevant simple minimal model mass terms into a mass matrix. \vskip 5mm
\paragraph{\bf Sector $(6,1,-\frac{4}{3})\oplus (\bar 6,1,+\frac{4}{3})$, $T^{c}_{3}=0$, $T^{c}_{8}=\pm \frac{2}{\sqrt{3}}$ states:}
\be
m_\Sigma+\eta  (-a+p+2 \omega )
\ee
\begin{center}
\begin{tabular}{|c|c|c|c|}
\hline 12 & Pati-Salam origin & ($C^{c}\!\!, T^{c}_{3},  T^{c}_{8}, B-L, C^{L}\!\!,T^{L}_{3}, C^{R}\!\!,  T^{R}_{3}$) &  phase convention\\
\hline
$R_{1}$ & $(\overline{10},1,3)_{126}$ &  $(\frac{10}{3},0,+\frac{2}{\sqrt{3}},-\frac{2}{3},0,0,2,-1)_{{126}}$ & $ -i\Sigma[1, 3, 5, 7, 9] +\ldots$ \\
\hline
$C_{1}$ & $({10},1,3)_{\overline{126}}$&  $(\frac{10}{3},0,-\frac{2}{\sqrt{3}},+\frac{2}{3},0,0,2,+1)_{\overline{126}}$ & $+i\overline{\Sigma}[1, 3, 5, 7, 9] +\ldots$ \\ 
\hline \end{tabular}
\end{center}
\vskip 5mm
\paragraph{\bf Sector $(6,1,-\frac{1}{3})\oplus (\bar 6,1,+\frac{1}{3})$, $T^{c}_{3}=0$, $T^{c}_{8}=\pm \frac{2}{\sqrt{3}}$ states:}
\be
\left(
\begin{array}{cc}
 m_\Sigma+(p-a) \eta  & -2 \sqrt{2} \beta  \omega  \\
 -2 \sqrt{2} \bar\beta \omega  & 2 (m_\Psi-a \rho )
\end{array}
\right)
\ee
\begin{center}
\begin{tabular}{|c|c|c|c|}
\hline 24 & Pati-Salam origin & ($C^{c}\!\!, T^{c}_{3},  T^{c}_{8}, B-L, C^{L}\!\!,T^{L}_{3}, C^{R}\!\!,  T^{R}_{3}$) &  phase convention\\
\hline
$R_{1}$ &  $(\overline{10},1,3)_{126}$&  $(\frac{10}{3},0,+\frac{2}{\sqrt{3}},-\frac{2}{3},0,0,2,0)_{{126}}$ & $-i\Sigma[1, 2, 5, 7, 9] +\ldots$ \\
$R_{2}$ & $(\overline{10},1,1)_{120}$&  $(\frac{10}{3},0,+\frac{2}{\sqrt{3}},-\frac{2}{3},0,0,0,0)_{{120}}$ & $ \Psi[5,7,9]+\ldots$ \\
\hline
$C_{1}$ &  $({10},1,3)_{\overline{126}}$&  $(\frac{10}{3},0,-\frac{2}{\sqrt{3}},+\frac{2}{3},0,0,2,0)_{\overline{126}}$ & $+i\overline{\Sigma}[1, 2, 5, 7, 9] +\ldots$ \\ 
$C_{2}$ & $({10},1,1)_{120}$&  $(\frac{10}{3},0,-\frac{2}{\sqrt{3}},+\frac{2}{3},0,0,0,0)_{{120}}$ & $ \Psi[5,7,9] +\ldots$ \\
\hline \end{tabular}
\end{center}
\vskip 5mm
\paragraph{\bf Sector $(6,1,+\frac{2}{3})\oplus (\bar 6,1,-\frac{2}{3})$, $T^{c}_{3}=0$, $T^{c}_{8}=\pm \frac{2}{\sqrt{3}}$ states:}
\be
m_\Sigma+\eta  (-a+p-2 \omega )
\ee
\begin{center}
\begin{tabular}{|c|c|c|c|}
\hline 12 & Pati-Salam origin & ($C^{c}\!\!, T^{c}_{3},  T^{c}_{8}, B-L, C^{L}\!\!,T^{L}_{3}, C^{R}\!\!,  T^{R}_{3}$) &  phase convention\\
\hline
$R_{1}$ & $(\overline{10},1,3)_{126}$&  $(\frac{10}{3},0,+\frac{2}{\sqrt{3}},-\frac{2}{3},0,0,2,+1)_{{126}}$ & $-i\Sigma[1, 3, 5, 7, 9] +\ldots$ \\
\hline
$C_{1}$ & $({10},1,3)_{\overline{126}}$&  $(\frac{10}{3},0,-\frac{2}{\sqrt{3}},+\frac{2}{3},0,0,2,-1)_{\overline{126}}$ & $+i\overline{\Sigma}[1, 3, 5, 7, 9] +\ldots$ \\ 
\hline \end{tabular}
\end{center}
\vskip 5mm
\subsubsection{$C^{c}=\frac{10}{3},  C^{L}=\frac{3}{4}$ - colour sextets \& antisextets, $SU(2)_{L}$-doublets}
This subspace is identical to the relevant part of the MSGUT, see e.g. \cite{Bajc:2004xe} and references therein.
\vskip 5mm
\paragraph{\bf Sector $(6,2,-\frac{5}{6})\oplus (\bar 6,2,+\frac{5}{6})$, $T^{c}_{3}=0$, $T^{c}_{8}=\pm \frac{2}{\sqrt{3}}$ states:}
\be
2 (m_\Phi+\lambda  (\omega -a))
\ee
\begin{center}
\begin{tabular}{|c|c|c|c|}
\hline 24 & Pati-Salam origin & ($C^{c}\!\!, T^{c}_{3},  T^{c}_{8}, B-L, C^{L}\!\!,T^{L}_{3}, C^{R}\!\!,  T^{R}_{3}$) &  phase convention\\
\hline
$R_{1}$ & $(\overline{10},2,2)_{{210}}$ &  $(\frac{10}{3},0,+\frac{2}{\sqrt{3}},-\frac{2}{3},
   \frac{3}{4},+\frac{1}{2},\frac{3}{4},-\frac{1}{2})_{{210}}$ & $-i\Phi[3, 5, 7, 9] +\ldots$ \\
\hline
$C_{1}$ &$({10},2,2)_{{210}}$ &  $(\frac{10}{3},0,-\frac{2}{\sqrt{3}},+\frac{2}{3},
   \frac{3}{4},-\frac{1}{2},\frac{3}{4},+\frac{1}{2})_{{210}}$ & $+i\Phi[3, 5, 7, 9]+\ldots$ \\
\hline \end{tabular}
\end{center}
\vskip 5mm
\paragraph{\bf Sector $(6,2,+\frac{1}{6})\oplus (\bar 6,2,-\frac{1}{6})$, $T^{c}_{3}=0$, $T^{c}_{8}=\pm \frac{2}{\sqrt{3}}$ states:}
\be
2 (m_\Phi-\lambda  (\omega +a))
\ee
\begin{center}
\begin{tabular}{|c|c|c|c|}
\hline 24 & Pati-Salam origin & ($C^{c}\!\!, T^{c}_{3},  T^{c}_{8}, B-L, C^{L}\!\!,T^{L}_{3}, C^{R}\!\!,  T^{R}_{3}$) &  phase convention\\
\hline
$R_{1}$ & $(\overline{10},2,2)_{{210}}$ &  $(\frac{10}{3},0,+\frac{2}{\sqrt{3}},-\frac{2}{3},
   \frac{3}{4},+\frac{1}{2},\frac{3}{4},+\frac{1}{2})_{{210}}$ & $i\Phi[1, 5, 7, 9] +\ldots$ \\
\hline
$C_{1}$ & $({10},2,2)_{{210}}$ &  $(\frac{10}{3},0,-\frac{2}{\sqrt{3}},+\frac{2}{3},
   \frac{3}{4},-\frac{1}{2},\frac{3}{4},-\frac{1}{2})_{{210}}$ & $-i\Phi[1, 5, 7, 9]  +\ldots$ \\
\hline \end{tabular}
\end{center}
\vskip 5mm
\subsubsection{$C^{c}=\frac{10}{3},  C^{L}=2$ - colour sextets \& antisextets, $SU(2)_{L}$-triplets}
Finally, this sector is again identical to its minimal model counterpart. 
\vskip 5mm
\paragraph{\bf Sector $(6,3,-\frac{1}{3})\oplus (\bar 6,3,+\frac{1}{3})$, $T^{c}_{3}=0$, $T^{c}_{8}=\pm \frac{2}{\sqrt{3}}$ states:}
\be
m_\Sigma-(a+p) \eta 
\ee
\begin{center}
\begin{tabular}{|c|c|c|c|}
\hline 36 & Pati-Salam origin & ($C^{c}\!\!, T^{c}_{3},  T^{c}_{8}, B-L, C^{L}\!\!,T^{L}_{3}, C^{R}\!\!,  T^{R}_{3}$) &  phase convention\\
\hline
$R_{1}$ & $(\overline{10},3,1)_{\overline{126}}$&  $(\frac{10}{3},0,+\frac{2}{\sqrt{3}},-\frac{2}{3},
   2,0,0,0)_{\overline{126}}$ & $-i\overline{\Sigma}[1, 2, 5, 7, 9] +\ldots$ \\
\hline
$C_{1}$ & $({10},3,1)_{{126}}$&  $(\frac{10}{3},0,-\frac{2}{\sqrt{3}},+\frac{2}{3},
   2,0,0,0)_{{126}}$ & $+i\Sigma[1, 2, 5, 7, 9]+\ldots$ \\
\hline \end{tabular}
\end{center}
\vskip 5mm
To conclude, in this section we have written down the mass matrices for all the 592 bosonic degrees of freedom (up to gauge transformations) of the Higgs sector of the next-to-minimal supersymmetric SUSY SO(10) model.  
We have checked that in the $m_{\Psi}\to \infty$ limit (corresponding to the decoupling of the extra ${\bf 120}_{\rm H}$ multiplet) the minimal model effective particle content is recovered (i.e. exactly 120 states decouple) and the resulting mass matrices are reduced to the relevant MSGUT formulae. This provides a non-trivial consistency cross-check of the results given previously in \cite{Bajc:2004xe} as well as ours.    
\section{Consistency check - Goldstone bosons\label{goldstones}}
Another nontrivial consistency check of some of our results consists in identifying the would-be Goldstone bosons associated to the spontaneous breakdown of the $SO(10)$ symmetry to its SM subgroup $SU(3)_{c}\otimes SU(2)_{L}\otimes U(1)_{Y}$. Since the gauge bosons associated to the generators in the coset $SO(10)/SU(3)_{c}\otimes SU(2)_{L}\otimes U(1)_{Y}$ should all become massive, there should be in total 33 Godstone bosons with the quantum numbers of the coset gauge fields (i.e. $(3,2,-\frac{5}{6})\oplus (\bar 3,2,+\frac{5}{6})$, $(3,2,+\frac{1}{6})\oplus (\bar 3,2,-\frac{1}{6})$, $(3,1,+\frac{2}{3})\oplus (\bar 3,1,-\frac{2}{3})$, $(1,1,-1)\oplus (1,1,+1)$ and $(1,1,0)$ sectors respectively) providing the relevant longitudinal components. The situation in the minimal model has been thoroughly studied in \cite{Bajc:2004xe} and the zeros in the Higgs spectra were revealed. 

With an extra ${\bf 120}_{\rm H}$ in the Higgs sector there are extra components contributing to the $(3,2,+\frac{1}{6})\oplus (\bar 3,2,-\frac{1}{6})$, $(3,1,+\frac{2}{3})\oplus (\bar 3,1,-\frac{2}{3})$ and  $(1,1,-1)\oplus (1,1,+1)$ sectors while the mass matrices of $(3,2,-\frac{5}{6})\oplus (\bar 3,2,+\frac{5}{6})$ and $(1,1,0)$ remain unaffected and the results of  \cite{Bajc:2004xe} are easily recovered. In what follows we shall focus on the former case and provide an evidence that the zeros are still in the spectra of the extended mass matrices, as required by consistency. 
\subsection{Relevant mass matrices in the Standard model vacuum}
Adopting the SM vacuum notation along the lines of \cite{Bajc:2004xe} (recall that the vacuum structure of the model with ${\bf 120}_{\rm H}$ is identical to the one of the MSGUT because there is no extra full SM singlet in there; c.f. also formulae (\ref{SMvacuumparameters}) and (\ref{sses}) in Section \ref{breakdown})
the relevant mass matrices can be recast in terms of the microscopic parameters as follows:
\vskip 5mm
\paragraph{\bf Sector $(\bar 3,1,-\frac{2}{3})\oplus (3,1,+\frac{2}{3})$:}
\be
2m_{\Phi}\left(
\begin{array}{cccc}
 \frac{\left(x^2-4 x+1\right) \eta }{\lambda -x \lambda } & \frac{\overline{s} \eta }{\sqrt{2}
   \sqrt{\eta  \lambda }} & -\frac{\overline{s} \eta }{\sqrt{\eta  \lambda }} & -\frac{\sqrt{2}
   \left(x^2-4 x+1\right) \beta }{(x-1) \lambda } \\
 \frac{s \eta }{\sqrt{2} \sqrt{\eta  \lambda }} & -\frac{x (x+1)}{x-1} & -\sqrt{2} x & \frac{s
   \beta }{\sqrt{\eta  \lambda }} \\
 -\frac{s \eta }{\sqrt{\eta  \lambda }} & -\sqrt{2} x & \frac{4 x^3}{(x-1)^2} & -\frac{\sqrt{2}
   s \beta }{\sqrt{\eta  \lambda }} \\
 -\frac{\sqrt{2} \left(x^2-4 x+1\right) \overline{\beta}}{(x-1) \lambda } & \frac{\overline{s}
   \overline{\beta}}{\sqrt{\eta  \lambda }} & -\frac{\sqrt{2} \overline{s} \overline{\beta}}{\sqrt{\eta  \lambda }} & r+\frac{x \left(7 x^2-4 x+1\right) \rho
   }{(x-1)^2 \lambda }
\end{array}
\right)
\ee
\vskip 5mm
\paragraph{\bf Sector $(\bar 3,2,-\frac{1}{6})\oplus (3,2,+\frac{1}{6})$ :}
\be
2m_{\Phi}\left(
\begin{array}{ccccc}
 \frac{\left(-4 x^3+6 x^2-5 x+1\right) \eta }{(x-1)^2 \lambda } & 0 & 0 & 0 & \frac{\sqrt{2}
   (1-2 x)^2 (x+1) \beta }{(x-1)^2 \lambda } \\
 0 & \frac{\left(-3 x^3+4 x^2-4 x+1\right) \eta }{(x-1)^2 \lambda } & -\frac{s\eta }{
   \sqrt{2\eta  \lambda }} & \frac{s\eta }{\sqrt{\eta  \lambda }} & -\frac{\sqrt{2} \left(3
   x^3-4 x^2+4 x-1\right) \overline{\beta}}{(x-1)^2 \lambda } \\
 0 & -\frac{\overline{s}\eta }{\sqrt{2\eta  \lambda }} & x+1 & -\sqrt{2} x &
   -\frac{\overline{s}\overline{\beta}}{\sqrt{\eta  \lambda }} \\
 0 & \frac{\overline{s}\eta }{\sqrt{\eta  \lambda }} & -\sqrt{2} x & -\frac{2 x}{x-1} &
   \frac{\sqrt{2} \overline{s}\overline{\beta}}{\sqrt{\eta  \lambda }} \\
 \frac{\sqrt{2} (1-2 x)^2 (x+1) \overline{\beta}}{(x-1)^2 \lambda } & -\frac{\sqrt{2} \left(3
   x^3-4 x^2+4 x-1\right) \beta }{(x-1)^2 \lambda } & -\frac{s\beta }{\sqrt{\eta  \lambda }} &
   \frac{\sqrt{2} s\beta }{\sqrt{\eta  \lambda }} & r+\frac{\left(x^2-4
   x+1\right) \rho }{(x-1) \lambda }
\end{array}
\right)\ee
\vskip 5mm
\paragraph{\bf Sector $(1,1,-1)\oplus (1,1,+1)$:}
\be
2m_{\Phi}\left(
\begin{array}{ccc}
 -\frac{3 x \eta }{\lambda } & \sqrt{\frac{3}{2}}\frac{ s \eta }{\sqrt{\eta  \lambda }} &
   -\frac{3 \sqrt{2} x \overline{\beta}}{\lambda } \\
 \sqrt{\frac{3}{2}}\frac{ \overline{s} \eta }{\sqrt{\eta  \lambda }} & \frac{3 x^3-x^2+3
   x-1}{(x-1)^2} & \frac{\sqrt{3} \overline{s} \overline{\beta}}{\sqrt{\eta  \lambda }} \\
 -\frac{3 \sqrt{2} x \beta }{\lambda } & \frac{\sqrt{3} s \beta }{\sqrt{\eta  \lambda }} &
   r-\frac{3 \left(x^2+2 x-1\right) \rho }{(x-1) \lambda }
\end{array}
\right)
\ee
It is not hard to see that with (\ref{sses}) all these mass matrices have indeed a zero in the spectra irrespective of the values of $r\equiv m_{\Psi}/m_{\Phi}$, $\beta$ and $\overline{\beta}$.
\section{The NMSGUT Yukawa sector\label{sect-yukawasector}}
The Yukawa structure of the theory provides an important and (at least in principle) testable imprint of the GUT-scale physics on the electroweak scale observables. In this section, we shall discuss some of the salient features of this part of the NMSGUT model.
 With an extra $120_{\rm H}$ at hand, the MSGUT \cite{Aulakh:2003kg} Yukawa piece of the superpotential
\be\label{WYmin}
W_{\rm Y}^{\mathrm{min}}=f_{10}^{ij}{\bf 16_{\rm F}^{\it i}}.{\bf 16_{\rm F}^{\it j}}.{\bf 10_{\rm H}}+f_{ \overline{126}}^{ij}{\bf 16_{\rm F}^{\it i}}.{\bf 16_{\rm F}^{\it j}}.{\bf \overline{126}_{\rm H}}
\ee
is extended into
\be\label{extendedYukawagrouptheory}
W_{\rm Y}=f_{10}^{ij}{\bf 16_{\rm F}^{\it i}}.{\bf 16_{\rm F}^{\it j}}.{\bf 10_{\rm H}}+f_{\overline{126}}^{ij}{\bf 16_{\rm F}^{\it i}}.{\bf 16_{\rm F}^{\it j}}.{\bf \overline{126}_{\rm H}}+f_{120}^{ij}{\bf 16_{\rm F}^{\it i}}.{\bf 16_{\rm F}^{\it j}}.{\bf {120}_{\rm H}}
\ee
where $f_{10}$, $f_{\overline{126}}$ are general complex $3\times 3$ matrices, the former two being symmetric in the family space while $f_{120}$ is antisymmetric.
After the $SO(10)$ breakdown, this structure gives rise to all the Yukawa interactions of the effective theory, in particular those of the $SU(2)_{L}$ doublets governing the effective matter sector mass sum-rules, the interactions of the coloured triplets entering e.g. the $d=5$ proton decay operators etc. Since the $d=5$ proton decay is rather elusive in SUSY GUTs unless the soft-SUSY spectra become specified let us in what follows focus namely on the former, i.e. the effective sum-rules for the Dirac and Majorana masses of the MSSM matter fermions. 
\subsection{Effective Yukawa sector sum-rules}
When the light doublet components in $\bf 10_{\rm H}$, $\bf \overline{126}_{\rm H}$ and $\bf 120_{\rm H}$   receive their electroweak VEVs, $W_{\rm Y}$ gives rise to the set of sum-rules  for the effective Yukawa couplings $Y_{u}$, $Y_{d}$, $Y_{e}$, $Y_{\nu}$ etc. of the form:
\bea
M_{u}\equiv Y_{u}v_{u}&=&Y_{10}v_{10}^{u}+Y_{126}v_{\overline{126}}^{u}+Y_{120}\left[v_{120}^{u(1)}+v_{120}^{u(2)}\right]\nn\\
M_{\nu}^{D}\equiv Y_{\nu}v_{u}&=&Y_{10}v_{10}^{u}-3Y_{126}v_{\overline{126}}^{u}+Y_{120}\left[v_{120}^{u(1)}-3v_{120}^{u(2)}\right]\label{NMSGUTsumrules}\\
M_{d}\equiv Y_{d}v_{d}&=&Y_{10}v_{10}^{d}+Y_{126}v_{\overline{126}}^{d}+Y_{120}\left[v_{120}^{d(1)}+v_{120}^{d(2)}\right]\nn\\
M_{e}\equiv Y_{e}v_{d}&=&Y_{10}v_{10}^{d}-3Y_{126}v_{\overline{126}}^{d}+Y_{120}\left[v_{120}^{d(1)}-3v_{120}^{d(2)}\right]\nn\\
M_{\nu}^{R}&=& Y_{126}V_{R}\nn\\
M_{\nu}^{L}&=& Y_{126}v_{L}\nn
\eea
where $Y_{10}$, $Y_{126}$ and  $Y_{120}$ are matrices proportional to $f_{10}$,  $f_{\overline{126}}$ and $f_{120}$ respectively (the exact ``matching'' formulae for the relevant proportionality factors in terms of the microscopic parameters of the model are the very subject of the next section) and the various $v_{\bf R}^{u,d}$ factors correspond to the projections of the electroweak VEVs onto the directions of the various $SU(2)_{L}$ doublets (residing in ${\bf R}\equiv {\bf 10}_{\rm H}$, ${\bf \overline{126}}_{\rm H}$ and ${\bf  {120}}_{\rm H}$) in the defining basis. Apart from that, $v_{L}$ and $V_{R}$ are the $B-L$ breaking VEVs of the $SU(2)_{L}$ triplet and singlet respectively, which are responsible for the Majorana masses of neutrinos. Ignoring the details of the underlying theory, the $v_{\bf R}^{u,d}$ factors are essentially arbitrary and one can naturally expect that the system (\ref{NMSGUTsumrules}) should admit good fits of all the low-energy matter fermion masses and mixings.

If, on the other hand, the Higgs sector of the model is fully specified, the weights above become computable in terms of the parameters entering the Higgs scalar potential (\ref{Wmin}), (\ref{W120}) and the VEVs driving the breakdown of the GUT symmetry (\ref{VEVs}). In case of the minimal scenario this has been done in full generality by Bajc, Melfo, Senjanovi\'c and Vissani in their 2004 paper  \cite{Bajc:2004xe} and further extended in \cite{Bajc:2005qe}. Subsequently, it has been pointed out \cite{Aulakh:2005bd,absoluteneutrinomassscale,Aulakh:2006vi} that the minimal setting is indeed incompatible with the low-energy data.

\subsection{Microscopic structure of the effective MSSM mass sum-rules in NMSGUT}
In this section we consider the extended Yukawa sector (\ref{extendedYukawagrouptheory}) and compute the effective projections of the electroweak VEVs entering (\ref{NMSGUTsumrules}) by means of the microscopic parameters of the theory. 
\subsubsection{Bidoublet mass matrix in NMSGUT}
In order to do that we should look at the shape of the $SU(2)_{L}$ doublet mixing arising from the generic 6$\times$6 bidoublet mass matrix (\ref{doubletmassmatrix}) emerging after spontaneous breakdown of the GUT symmetries. 
Notice that the upper-left 4$\times$4 sub-block of (\ref{doubletmassmatrix}) corresponding to the doublets in $\bf 10_{\rm H}$, $\bf \overline{126}_{\rm H}\oplus 126_{\rm H}$, $\bf 210_{\rm H}$  of the original MSGUT is indeed identical (up to an irrelevant global rephasing) to the bidoublet mass matrix given in  \cite{Bajc:2004xe}, which provides a non-trivial consistency check of both ours and Bajc \& co.'s analysis. The last two rows/columns in (\ref{doubletmassmatrix}) are due to the pair of bidoublets in $\bf 120_{\rm H}$ and we have chosen our convention in such a way there are no pending $i$ factors.

Adopting the SM vacuum notation along the lines of \cite{Bajc:2004xe}
the VEVs in the mass matrix under consideration can be recast in terms of the microscopic parameters (\ref{SMvacuumparameters}) as:
\bea
\label{doubletmassmatrixinSMvacuum}
& & M_{(1,2,\pm 1)}\propto \begin{pmatrix}
2 \frac{m_{H}}{m_{\Phi}} & 
\frac{\overline\alpha}{\lambda}\frac{3}{2}\frac{3x-1}{x-1}&
-\frac{\alpha}{\lambda}\frac{3}{2}\frac{2x^{2}+x-1}{x-1}&
-\overline{\alpha}\overline{\sigma}&
\frac{\gamma}{\lambda}\frac{x(5x^{2}-1)}{(x-1)^{2}} &
-\frac{\gamma}{\lambda}\sqrt{3}x\\
\frac{\alpha}{\lambda}\frac{3}{2}\frac{3x-1}{x-1}&
\frac{\eta}{\lambda}\frac{-8 x^3+9 x^2-6 x+1}{(x-1)^2}&
0&
\sqrt{6}\eta \overline{\sigma}&
-\frac{\beta}{\lambda}\sqrt{6} x &
\frac{\beta}{\lambda}\sqrt{2}\frac{x \left(7 x^2-4 x+1\right)}{(x-1)^2}\\
-\frac{\overline\alpha}{\lambda}\frac{3}{2}\frac{2x^{2}+x-1}{x-1}&
0&
\frac{\eta}{\lambda}\frac{-12 x^3+17 x^2-10 x+1}{(x-1)^2}&
0&
-\frac{\overline{\beta}}{\lambda}\sqrt{6}x&
\frac{\bar\beta}{\lambda}\sqrt{2}\frac{x \left(3 x^2+4 x-3\right)}{(x-1)^2}\\
-\alpha\sigma&
\sqrt{6}\eta \sigma&
0&
-4\frac{4x-1}{x-1}&
2\beta\sigma&
-2\sqrt{3}\beta\sigma\\
\frac{\gamma}{\lambda}\frac{x(5x^{2}-1)}{(x-1)^{2}} &
-\frac{\bar\beta}{\lambda}\sqrt{6} x &
-\frac{\beta}{\lambda}\sqrt{6} x &
2\overline{\beta}\overline{\sigma}&
2\frac{m_{\Psi}}{m_{\Phi}}&
-2\sqrt{3}\frac{\rho}{\lambda}x \\
-\frac{\gamma}{\lambda}\sqrt{3}x&
\frac{\bar\beta}{\lambda}\sqrt{2}\frac{x \left(7 x^2-4 x+1\right)}{(x-1)^2}&
\frac{\beta}{\lambda}\sqrt{2}\frac{x \left(3 x^2+4 x-3\right)}{(x-1)^2}&
-2\sqrt{3}\overline{\beta}\overline{\sigma}&
-2\sqrt{3}\frac{\rho}{\lambda}x&
2\frac{m_{\Psi}}{m_{\Phi}}-4\frac{\rho}{\lambda}\frac{\left(x^2+2 x-1\right)}{x-1}
\end{pmatrix}\nn
\eea
where now also $\sigma$ and $\overline\sigma$ are functions of the basic parameters and obey (c.f. \cite{Bajc:2004xe}), c.f. (\ref{sses}).
\subsubsection{Arranging the light MSSM Higgs doublets}
As in the case of the minimal model, $m_{H}$ can be fixed from the zero-determinant condition which is necessary to arrange the pair of light MSSM-like Higgs doublets $h_{u}$ and $h_{d}$. In the mass basis, these are also the only zero modes of this mass matrix while the orthogonal states correspond to the five heavy (typically GUT-scale) Higgs doublets $H_{u,d}^{(1)}, \ldots, H_{u,d}^{(5)}$.

Up to an overall normalization, the light Higgs doublets of the MSSM correspond to the following combinations of the defining basis states (using $H^{u,d}$ for the doublets within $(1,2,2)_{10}$ and so on, i.e. $\Sigma^{u,d}\in (15,2,2)_{126}$,  $\overline{\Sigma}^{u,d}\in (15,2,2)_{\overline{126}}$, $\Phi^{u}\in (10,2,2)_{210}$, $\Phi^{d}\in (\overline{10},2,2)_{210}$, $\Psi^{u,d}_{(1)}\in (1,2,2)_{120}$ and $\Psi_{(2)}^{u,d}\in (15,2,2)_{120}$): 
\bea
h_{u}&\propto& w^{u}_{10}H^{u}+w^{u}_{\overline{126}}\overline{\Sigma}^{u}+w^{u}_{{126}}{\Sigma}^{u}+w^{u}_{210}\Phi^{u}+w^{u(1)}_{120}\Psi^{u}_{(1)}+w^{u(2)}_{120}\Psi^{u}_{(2)}\nn\\
h_{d}&\propto& w^{d}_{10}H^{d}+w^{d}_{{126}}{\Sigma}^{d}+w^{d}_{\overline{126}}\overline{\Sigma}^{d}+w^{d}_{210}\Phi^{d}+w^{d(1)}_{120}\Psi^{d}_{(1)}+w^{d(2)}_{120}\Psi^{d}_{(2)}
\label{doubletweights}
\eea
where the generic $w_{\bf R}^{u,d}$ factors stand for the numerical weights (or projections) of the various components of $h_{u,d}$ in the relevant 6-dimensional SM-doublet space. Recall that in order for the change of basis (\ref{doubletweights}) to be unitary, the weights above should obey the normalization condition $\sum_{\bf R}|w_{\bf R}^{u,d}|^{2}=1$ for each of the hypercharges.
\subsubsection{Projecting the MSSM Higgs doublets onto the defining MSGUT basis}
Defining a ``decoupling parameter''
$r\equiv {m_{\Phi} / m_{\Psi}}$ which is a quantity that can trace the ``strength of interaction'' between the minimal model bidoublets residing in $\bf 10_{\rm H}$, $\bf \overline{126}_{\rm H}\oplus 126_{\rm H}$ and $\bf 210_{\rm H}$, and the extra bidoublets within $\bf 120_{\rm H}$. (indeed, for $m_{\Psi}\to \infty$ we get $r\to 0$ and the minimal model situation should be recovered) the relevant weight factors can be shown 
to obey the following formulae (up to an overall normalization):
\bea
%
w_{10}^{u*}&\equiv& 2\left[
P_{11}^{a}+
r\left(\frac{\beta \bar\beta}{\eta\lambda}P_{12}^{a} +\frac{\rho}{\lambda}P_{12}^{b}\right)+
r^{2}\left( \frac{\beta^2 \bar\beta^2}{\eta^{2}\lambda^{2}}P_{13}^{a}+\frac{\rho\beta\bar\beta}{\eta\lambda^{2}} P_{13}^{b} +
   \frac{\rho^{2}}{\lambda^{2}} P_{13}^{c}\right)\right]\nn\\
w_{\overline{126}}^{u*}&\equiv& 
\sqrt{\frac{2}{3}}\left[
\frac{ \alpha}{\eta }P_{11}^{b} +r \left(\frac{\bar\alpha \beta ^2}{\eta^{2}\lambda }P_{12}^{c}+\frac{\alpha  \bar\beta \beta}{\eta^{2}\lambda }P_{12}^{d} 
+\frac{ \beta\gamma }{\eta\lambda }P_{12}^{e} +\frac{\alpha\rho}{\eta\lambda }P_{12}^{f}  \right)+\right.\nn\\
& &\qquad +\left.
r^2 \left(
\frac{\alpha  \bar\beta^2 \beta ^2}{\eta^{3}\lambda^{2} }P_{12}^{g} +
\frac{\bar\alpha \bar\beta \beta ^3}{\eta^{3}\lambda^{2} }P_{12}^{h} +
\frac{\bar\beta \beta ^2\gamma }{\eta^{2}\lambda^{2} }P_{13}^{d}+  
\frac{ \bar\alpha\beta ^2\rho}{\eta^{2}\lambda^{2} }P_{13}^{e}+
\frac{\alpha  \bar\beta \beta\rho}{\eta^{2}\lambda^{2} }P_{13}^{f}+
\frac{\beta\gamma\rho}{\eta\lambda^{2} }P_{13}^{g}+
\frac{  \alpha  \rho^{2}}{\eta\lambda^{2} }P_{13}^{h}
\right)      
\right]
\nn\\
w_{126}^{u*}&\equiv& 
\sqrt{\frac{2}{3}}\left[
\frac{ \bar\alpha}{\eta }P_{11}^{c} +r \left(\frac{\alpha \bar\beta ^2}{\eta^{2}\lambda }P_{12}^{i}+\frac{\bar\alpha  \beta \bar\beta}{\eta^{2}\lambda }P_{12}^{j} 
+\frac{ \bar\beta\gamma }{\eta\lambda }P_{12}^{k} +\frac{\bar\alpha\rho}{\eta\lambda }P_{12}^{l}  \right)+\right.\nn\\
& &\qquad +\left.
r^2 \left(
\frac{\alpha  \bar\beta^2\beta ^2}{\eta^{3}\lambda^{2} }P_{13}^{i} +
\frac{\alpha\beta \bar\beta ^3}{\eta^{3}\lambda^{2} }P_{13}^{j} +
\frac{\beta \bar\beta ^2\gamma }{\eta^{2}\lambda^{2} }P_{13}^{k}+  
\frac{\alpha\bar\beta ^2\rho}{\eta^{2}\lambda^{2} }P_{13}^{l}+
\frac{\bar\alpha\beta \bar\beta\rho}{\eta^{2}\lambda^{2} }P_{13}^{m}+
\frac{\bar\beta\gamma\rho}{\eta\lambda^{2} }P_{13}^{n}+
\frac{\bar\alpha  \rho^{2}}{\eta\lambda^{2} }P_{13}^{o}
  \right)      
\right]
\nn
\\
w_{210}^{u*}&\equiv& 
\frac{\bar\sigma}{\omega}x\left[
\frac{ \alpha}{\lambda }P_{10}^{a} +r \left(
\frac{\bar\alpha \beta ^2}{\eta\lambda^{2} }P_{11}^{d}+
\frac{\alpha  \bar\beta \beta}{\eta\lambda^{2} }P_{11}^{e} +
\frac{ \beta\gamma }{\lambda^{2} }P_{11}^{f} +
\frac{\alpha\rho}{\lambda^{2} }P_{11}^{g}  
\right)+\right.\nn\\
& &\qquad +\left.
r^2 \left(
\frac{\alpha  \bar\beta^2 \beta ^2}{\eta^{2}\lambda^{3} }P_{12}^{m} +
\frac{\bar\alpha \bar\beta \beta ^3}{\eta^{2}\lambda^{3} }P_{12}^{n} +
\frac{\bar\beta \beta ^2\gamma }{\eta\lambda^{3} }P_{12}^{o}+  
\frac{ \bar\alpha\beta ^2\rho}{\eta\lambda^{3} }P_{12}^{p}+
\frac{\alpha  \bar\beta \beta\rho}{\eta\lambda^{3} }P_{12}^{q}+
\frac{\beta\gamma\rho}{\lambda^{3} }P_{12}^{r}+
\frac{  \alpha  \rho^{2}}{\lambda^{3} }P_{12}^{s}
\right)      
\right]
\nn\\
w_{120}^{u(1)*}&\equiv& 
\frac{1}{x-1}
\left[r\left(
\frac{\alpha\bar\beta}{\eta\lambda}P_{13}^{p}+
\frac{\bar\alpha\beta}{\eta\lambda}P_{13}^{q}+
\frac{\gamma}{\lambda}P_{13}^{r}
\right)+\right.\nn\\
& & \qquad+
\left. r^{2}\left(
\frac{\alpha\beta\bar\beta^{2}}{\eta^{2}\lambda^{2}}P_{14}^{a}+
\frac{\bar\alpha\bar\beta\beta^{2}}{\eta^{2}\lambda^{2}}P_{14}^{b}+
\frac{\beta\bar\beta\gamma}{\eta\lambda^{2}}P_{14}^{c}+
\frac{\alpha\bar\beta\rho}{\eta\lambda^{2}}P_{14}^{d}+
\frac{\bar\alpha\beta\rho}{\eta\lambda^{2}}P_{14}^{e}+
\frac{\gamma\rho}{\lambda^{2}}P_{14}^{f}
\right)\right]\nn\\
w_{120}^{u(2)*}&\equiv& 
\frac{1}{\sqrt{3}}
\left[r\left(
\frac{\alpha\bar\beta}{\eta\lambda}P_{12}^{t}+
\frac{\bar\alpha\beta}{\eta\lambda}P_{12}^{u}+
\frac{\gamma}{\lambda}P_{12}^{v}
\right)+\right.\nn\\
& & \qquad+
\left. r^{2}\left(
\frac{\alpha\beta\bar\beta^{2}}{\eta^{2}\lambda^{2}}P_{13}^{s}+
\frac{\bar\alpha\bar\beta\beta^{2}}{\eta^{2}\lambda^{2}}P_{13}^{t}+
\frac{\beta\bar\beta\gamma}{\eta\lambda^{2}}P_{13}^{u}+
\frac{\alpha\bar\beta\rho}{\eta\lambda^{2}}P_{13}^{v}+
\frac{\bar\alpha\beta\rho}{\eta\lambda^{2}}P_{13}^{o}+
\frac{\gamma\rho}{\lambda^{2}}P_{13}^{w}
\right)\right]
\label{upsectorweights}
\eea
and similarly in the down-sector:
\bea
w_{10}^{d*}&\equiv& 2\left[
P_{11}^{a}+
r\left(\frac{\beta \bar\beta}{\eta\lambda}P_{12}^{a} +\frac{\rho}{\lambda}P_{12}^{b}\right)+
r^{2}\left( \frac{\beta^2 \bar\beta^2}{\eta^{2}\lambda^{2}}P_{13}^{a}+\frac{\rho\beta\bar\beta}{\eta\lambda^{2}} P_{13}^{b} +
\frac{\rho^{2}}{\lambda^{2}} P_{13}^{c}\right)\right]\nn\\
w_{126}^{d*}&\equiv& 
\sqrt{\frac{2}{3}}\left[
\frac{ \bar\alpha}{\eta }P_{11}^{b} +r \left(\frac{\alpha \bar\beta ^2}{\eta^{2}\lambda }P_{12}^{c}+\frac{\bar\alpha\beta\bar\beta}{\eta^{2}\lambda }P_{12}^{d} 
+\frac{\bar\beta\gamma }{\eta\lambda }P_{12}^{e} +\frac{\bar\alpha\rho}{\eta\lambda }P_{12}^{f}  \right)+\right.\nn\\
& &\qquad +\left.
r^2 \left(
\frac{\bar\alpha\beta^2 \bar\beta ^2}{\eta^{3}\lambda^{2} }P_{12}^{g} +
\frac{\alpha \beta \bar\beta ^3}{\eta^{3}\lambda^{2} }P_{12}^{h} +
\frac{\beta \bar\beta ^2\gamma }{\eta^{2}\lambda^{2} }P_{13}^{d}+  
\frac{\alpha\bar\beta ^2\rho}{\eta^{2}\lambda^{2} }P_{13}^{e}+
\frac{\bar\alpha\beta \bar\beta\rho}{\eta^{2}\lambda^{2} }P_{13}^{f}+
\frac{\bar\beta\gamma\rho}{\eta\lambda^{2} }P_{13}^{g}+
\frac{\bar\alpha  \rho^{2}}{\eta\lambda^{2} }P_{13}^{h}
\right)      
\right]
\nn
\\
w_{\overline{126}}^{d*}&\equiv& 
\sqrt{\frac{2}{3}}\left[
\frac{\alpha}{\eta }P_{11}^{c} +r \left(
\frac{\bar\alpha\beta ^2}{\eta^{2}\lambda }P_{12}^{i}+
\frac{\alpha\bar\beta\beta}{\eta^{2}\lambda }P_{12}^{j}+ 
\frac{\beta\gamma }{\eta\lambda }P_{12}^{k} +
\frac{\alpha\rho}{\eta\lambda }P_{12}^{l}  
\right)+\right.\nn\\
& &\qquad +\left.
r^2 \left(
\frac{\bar\alpha\beta^2\bar\beta ^2}{\eta^{3}\lambda^{2} }P_{13}^{i} +
\frac{\bar\alpha\bar\beta\beta ^3}{\eta^{3}\lambda^{2} }P_{13}^{j} +
\frac{\bar\beta\beta ^2\gamma }{\eta^{2}\lambda^{2} }P_{13}^{k}+  
\frac{\bar\alpha\beta ^2\rho}{\eta^{2}\lambda^{2} }P_{13}^{l}+
\frac{\alpha\bar\beta\beta\rho}{\eta^{2}\lambda^{2} }P_{13}^{m}+
\frac{\beta\gamma\rho}{\eta\lambda^{2} }P_{13}^{n}+
\frac{\alpha  \rho^{2}}{\eta\lambda^{2} }P_{13}^{o}
  \right)      
\right]
\nn
\\
w_{210}^{d*}&\equiv& 
\frac{\sigma}{\omega}x\left[
\frac{\bar\alpha}{\lambda }P_{10}^{a} +r \left(
\frac{\alpha\bar\beta ^2}{\eta\lambda^{2} }P_{11}^{d}+
\frac{\bar\alpha\beta\bar\beta}{\eta\lambda^{2} }P_{11}^{e} +
\frac{ \bar\beta\gamma }{\lambda^{2} }P_{11}^{f} +
\frac{\bar\alpha\rho}{\lambda^{2} }P_{11}^{g}  
\right)+\right.\nn\\
& &\qquad +\left.
r^2 \left(
\frac{\bar\alpha \beta^2 \bar\beta ^2}{\eta^{2}\lambda^{3} }P_{12}^{m} +
\frac{\alpha \beta \bar\beta ^3}{\eta^{2}\lambda^{3} }P_{12}^{n} +
\frac{\beta \bar\beta ^2\gamma }{\eta\lambda^{3} }P_{12}^{o}+  
\frac{\alpha\bar\beta ^2\rho}{\eta\lambda^{3} }P_{12}^{p}+
\frac{\bar\alpha  \beta \bar\beta\rho}{\eta\lambda^{3} }P_{12}^{q}+
\frac{\bar\beta\gamma\rho}{\lambda^{3} }P_{12}^{r}+
\frac{ \bar\alpha  \rho^{2}}{\lambda^{3} }P_{12}^{s}
\right)      
\right]
\nn
\\
w_{120}^{d(1)*}&\equiv& 
\frac{1}{x-1}
\left[r\left(
\frac{\bar\alpha\beta}{\eta\lambda}P_{13}^{p}+
\frac{\alpha\bar\beta}{\eta\lambda}P_{13}^{q}+
\frac{\gamma}{\lambda}P_{13}^{r}
\right)+\right.\nn\\
& & \qquad+
\left. r^{2}\left(
\frac{\bar\alpha\bar\beta\beta^{2}}{\eta^{2}\lambda^{2}}P_{14}^{a}+
\frac{\alpha\beta\bar\beta^{2}}{\eta^{2}\lambda^{2}}P_{14}^{b}+
\frac{\bar\beta\beta\gamma}{\eta\lambda^{2}}P_{14}^{c}+
\frac{\bar\alpha\beta\rho}{\eta\lambda^{2}}P_{14}^{d}+
\frac{\alpha\bar\beta\rho}{\eta\lambda^{2}}P_{14}^{e}+
\frac{\gamma\rho}{\lambda^{2}}P_{14}^{f}
\right)\right]\nn\\
w_{120}^{d(2)*}&\equiv& 
\frac{1}{\sqrt{3}}
\left[r\left(
\frac{\bar\alpha\beta}{\eta\lambda}P_{12}^{t}+
\frac{\alpha\bar\beta}{\eta\lambda}P_{12}^{u}+
\frac{\gamma}{\lambda}P_{12}^{v}
\right)+\right.\nn\\
& &\qquad +
\left. r^{2}\left(
\frac{\bar\alpha\bar\beta\beta^{2}}{\eta^{2}\lambda^{2}}P_{13}^{s}+
\frac{\alpha\beta\bar\beta^{2}}{\eta^{2}\lambda^{2}}P_{13}^{t}+
\frac{\bar\beta\beta\gamma}{\eta\lambda^{2}}P_{13}^{u}+
\frac{\bar\alpha\beta\rho}{\eta\lambda^{2}}P_{13}^{v}+
\frac{\alpha\bar\beta\rho}{\eta\lambda^{2}}P_{13}^{o}+
\frac{\gamma\rho}{\lambda^{2}}P_{13}^{w}
\right)\right]
\label{downsectorweights}
\eea
where the $P_{n}^{\rm x}$ factors denote polynomials in $x$ of order $n$ that are given (in terms of their $\mathbbm{Z}$-irreducible components) in Tables \ref{TableOfBigPolynomials} and \ref{TableOfIrreduciblePolynomials}.

\begin{table}[t]
\begin{tabular}{|c|l|}
\hline
$P_{10}^{a}$ & $-Q_{3}^{b} Q_{3}^{c} (x-1)^4$\\
\hline
$P_{11}^{a}$ & $-Q_{3}^{b} Q_{5}^{a} (x-1)^3$\\
$P_{11}^{b}$ & $-3 Q_{3}^{a}(x) Q_{3}^{b}(x) (x-1)^4 (3 x-1)$\\
$P_{11}^{c}$ & $3 Q_{5a}(x-1)^4 (x+1) (2 x-1)$\\
$P_{11}^{d}$ & $-3 Q_{2}^{b} (x-1)^4 x (x+1)^2 (2 x-1)^2$\\
$P_{11}^{e}$ & $-Q_{8}^{a} (x-1)^2 x$\\
$P_{11}^{f}$ & $-Q_{3}^{b} Q_{5}^{b} (x-1)^2 x$\\
$P_{11}^{g}$ & $2 Q_{2}^{a} Q_{3}^{b} Q_{3}^{c} (x-1)^3$\\
\hline
$P_{12}^{a}$ & $-4 Q_{10}^{a}(x-1) x$\\
$P_{12}^{b}$ & $2 Q_{2}^{a} Q_{3}^{b} Q_{5}^{a} (x-1)^2$\\
$P_{12}^{c}$ & $-6 Q_{3}^{d} (x-1)^2 x^2 (x+1)^2 (2 x-1)^2 (3 x-1)$\\
$P_{12}^{d}$ & $-6 Q_{8}^{b} (x-1)^2 x (3 x-1)$\\
$P_{12}^{e}$ & $-6 Q_{3}^{b} Q_{4}^{a} (x-1)^2 x^2 (3 x-1)$\\
$P_{12}^{f}$ & $6 Q_{2}^{a} Q_{3}^{a} Q_{3}^{b} (x-1)^3 (3 x-1)$\\
$P_{12}^{g}$ & $12 Q_{4}^{b} (x-1) x^3 (3 x-1) (5 x-3) \left(x^2+1\right)$\\
$P_{12}^{h}$ & $12 (x-1) x^3 (x+1)^2 (2 x-1)^2 (3 x-1) (5 x-3) \left(x^2+1\right)$\\
$P_{12}^{i}$ & $-12 Q_{5}^{c} (x-1)^2 x^2 (x+1) (2 x-1) (3 x-1)$\\
$P_{12}^{j}$ & $12 Q_{7}^{a} (x-1)^2 x (x+1) (2 x-1)$\\
$P_{12}^{k}$ & $-12 Q_{5}^{a} (x-1)^3 x^2 (x+1) (2 x-1)$\\
$P_{12}^{l}$ & $-6 Q_{2}^{a} Q_{5}^{a} (x-1)^3 (x+1) (2 x-1)$\\
$P_{12}^{m}$ & $-12 Q_{4}^{b} (x-1)^5 x^3$\\
$P_{12}^{n}$ & $-12 (x-1)^5 x^3 (x+1)^2 (2 x-1)^2$\\
$P_{12}^{o}$ & $-4 Q_{9}^{a} x^3$\\
$P_{12}^{p}$ & $3 Q_{4}^{c} (x-1)^3 x (x+1)^2 (2 x-1)^2$\\
$P_{12}^{q}$ & $3 Q_{8}^{c} (x-1)^3 x$\\
$P_{12}^{r}$ & $Q_{3}^{b} Q_{7}^{b} (x-1) x$\\
$P_{12}^{s}$ & $3 Q_{3}^{b} Q_{3}^{c} (x-1)^4 x^2$\\
$P_{12}^{t}$ & $3 Q_{3}^{b} Q_{5}^{d} (x-1)^2 x (3 x-1)$\\
$P_{12}^{u}$ & $3 Q_{2}^{c} Q_{5}^{a} (x-1)^2 x (x+1) (2 x-1)$\\
$P_{12}^{v}$ & $3 Q_{3}^{b} Q_{5}^{a} (x-1)^3 x$\\
\hline
\end{tabular}
\hskip 5mm
\begin{tabular}{|c|l|}
\hline
$P_{13}^{a}$ & $8 Q_{10}^{b} x^3$\\
$P_{13}^{b}$ & $2 Q_{11}^{a} (x-1) x$\\
$P_{13}^{c}$ & $3 Q_{3}^{b} Q_{5}^{a} (x-1)^3 x^2$\\
$P_{13}^{d}$ & $-12 Q_{9}^{c} x^3 (3 x-1)$\\
$P_{13}^{e}$ & $18 Q_{2}^{d} (x-1)^3 x^3 (x+1)^2 (2 x-1)^2 (3 x-1)$\\
$P_{13}^{f}$ & $6 Q_{9}^{b} (x-1)^2 x (3 x-1)$\\
$P_{13}^{g}$ & $6 Q_{3}^{b} Q_{5}^{e} (x-1) x^3 (3 x-1)$\\
$P_{13}^{h}$ & $9 Q_{3}^{a} Q_{3}^{b} (x-1)^4 x^2 (3 x-1)$\\
$P_{13}^{i}$ & $-12 (x-1) x^3 (x+1)^3 (2 x-1)^3 (3 x-1) \left(x^2+1\right)$\\
$P_{13}^{j}$ & $-12 Q_{4}^{b} (x-1) x^3 (x+1) (2 x-1) (3 x-1) \left(x^2+1\right)$\\
$P_{13}^{k}$ & $-12 Q_{7}^{c} (x-1) x^3 (x+1) (2 x-1)$\\
$P_{13}^{l}$ & $6 Q_{6}^{a} (x-1)^2 x^2 (x+1) (2 x-1) (3 x-1)$\\
$P_{13}^{m}$ & $-6 Q_{8}^{d} (x-1)^2 x (x+1) (2 x-1)$\\
$P_{13}^{n}$ & $6 Q_{2}^{e} Q_{5}^{a} (x-1) x^2 (x+1)^2 (2 x-1)$\\
$P_{13}^{o}$ & $-9 Q_{5}^{a} (x-1)^4 x^2 (x+1) (2 x-1)$\\
$P_{13}^{p}$ & $-Q_{3}^{b} Q_{5}^{f} (x-1)^3 x (3 x-1)$\\
$P_{13}^{q}$ & $-3 Q_{5}^{a} (x-1)^5 x (x+1) (2 x-1)$\\
$P_{13}^{r}$ & $-Q_{3}^{b} Q_{5}^{a} (x-1)^2 x \left(5 x^2-1\right)$\\
$P_{13}^{s}$ & $-6 Q_{9}^{d} (x-1) x^2 (3 x-1)$\\
$P_{13}^{t}$ & $-6 Q_{8}^{e} (x-1) x^2(x+1) (2 x-1)$\\
$P_{13}^{u}$ & $-6 Q_{10}^{c} (x-1) x^2$\\
$P_{13}^{v}$ & $-3 Q_{3}^{b} Q_{5}^{f} (x-1)^2 x^2 (3 x-1)$\\
$P_{13}^{w}$ & $-3 Q_{3}^{b} Q_{5}^{a} (x-1) x^2 \left(5 x^2-1\right)$\\
\hline
$P_{14}^{a}$ & $-2 Q_{10}^{d} (x-1) x^2 (3 x-1)$\\
$P_{14}^{b}$ & $-6 Q_{7}^{d} (x-1)^3 x^2 (x+1) (2 x-1)$\\
$P_{14}^{c}$ & $-2 Q_{12}^{a} x^2$\\
$P_{14}^{d}$ & $Q_{3}^{b} Q_{7}^{e} (x-1)^2 x (3 x-1)$\\
$P_{14}^{e}$ & $3 Q_{3}^{e} Q_{5}^{a} (x-1)^3 x (x+1) (2 x-1)$\\
$P_{14}^{f}$ & $Q_{3}^{b} Q_{4}^{d} Q_{5}^{a} (x-1) x$\\
\hline
\end{tabular}
\caption{\label{TableOfBigPolynomials} Table of the $P_{n}^{\rm x}(x)$ polynomials used in the text in terms of their $\mathbbm{Z}$-irreducible components}
\end{table}

\begin{table}
\begin{tabular}{|c|l|}
\hline
$Q_{2}^{a}$ & $x^2+2 x-1$\\
$Q_{2}^{b}$ & $2 x^2-5 x+1$\\
$Q_{2}^{c}$ & $3 x^2+4 x-3$\\
$Q_{2}^{d}$ & $2 x^2+3 x-3$\\
$Q_{2}^{e}$ & $7 x^2-6 x+1$\\
\hline
\end{tabular}
\hskip 1cm
\begin{tabular}{|c|l|}
\hline
$Q_{3}^{a}$ & $x^3+5 x-1$\\
$Q_{3}^{b}$ & $12 x^3-17 x^2+10 x-1$\\
$Q_{3}^{c}$ & $4 x^3-9 x^2+9 x-2$\\
$Q_{3}^{d}$ & $3 x^3+5 x^2+x-3$\\
$Q_{3}^{e}$ & $5 x^3+6 x^2-9 x+2$\\
\hline
\end{tabular}
\hskip 1cm
\begin{tabular}{|c|l|}
\hline
$Q_{4}^{a}$ & $4 x^4+5 x^3-5 x^2-x+1$\\
$Q_{4}^{b}$ & $2 x^4-6 x^3+21 x^2-16 x+3$\\
$Q_{4}^{c}$ & $7 x^4-5 x^3+6 x^2-5 x+1$\\
$Q_{4}^{d}$ & $13 x^4+11 x^3-3 x^2-7 x+2$\\
\hline
\end{tabular}
\vskip 5mm
\begin{tabular}{|c|l|}
\hline
$Q_{5}^{a}$ & $9 x^5+20 x^4-32 x^3+21 x^2-7 x+1$\\
$Q_{5}^{b}$ & $14 x^5-18 x^4+29 x^3-25 x^2+9 x-1$\\
$Q_{5}^{c}$ & $x^5-8 x^4-2 x^3-3 x^2+8 x-2$\\
$Q_{5}^{d}$ & $x^5-14 x^4-10 x^3+5 x^2+9 x-3$\\
$Q_{5}^{e}$ & $14 x^5+30 x^4-x^3-15 x^2-3 x+3$\\
$Q_{5}^{f}$ & $5 x^5-12 x^4+8 x^3+11 x^2-3 x-1$\\
\hline
\end{tabular}
\hskip 1cm
\begin{tabular}{|c|l|}
\hline
$Q_{6}^{a}$ & $7 x^6-23 x^5+10 x^4+47 x^3-38 x^2+4 x+1$\\
\hline
$Q_{7}^{a}$ & $12 x^7+36 x^6-74 x^5+64 x^4-44 x^3+25 x^2-8 x+1$\\
$Q_{7}^{b}$ & $40 x^7-37 x^6+69 x^5-39 x^4-2 x^3-3 x^2+5 x-1$\\
$Q_{7}^{c}$ & $18 x^7+4 x^6-147 x^5+181 x^4-114 x^3+60 x^2-21 x+3$\\
$Q_{7}^{d}$ & $18 x^7+11 x^6-142 x^5+157 x^4-102 x^3+61 x^2-22 x+3$\\
$Q_{7}^{e}$ & $13 x^7-49 x^6-30 x^5+123 x^4+34 x^3-72 x^2+11 x+2$\\
\hline
\end{tabular}
\vskip 5mm
\begin{tabular}{|c|l|}
\hline
$Q_{8}^{a}$ & $360 x^8-1342 x^7+2702 x^6-3417 x^5+2967 x^4-1660 x^3+528 x^2-77 x+3$\\
$Q_{8}^{b}$ & $54 x^8-170 x^7+387 x^6-441 x^5+474 x^4-398 x^3+183 x^2-35 x+2$\\
$Q_{8}^{c}$ & $128 x^8-548 x^7+767 x^6-341 x^5-211 x^4+286 x^3-115 x^2+19 x-1$\\
$Q_{8}^{d}$ & $12 x^8+44 x^7-245 x^6+195 x^5-33 x^4-10 x^3+9 x^2-5 x+1$\\
$Q_{8}^{e}$ & $18 x^8+21 x^7-91 x^6+261 x^5-307 x^4+191 x^3-81 x^2+23 x-3$\\
\hline
$Q_{9}^{a}$ & $210 x^9-128 x^8-173 x^7+127 x^6+337 x^5-599 x^4+405 x^3-135 x^2+21 x-1$\\
$Q_{9}^{b}$ & $84 x^9-303 x^8+500 x^7-529 x^6+222 x^5+118 x^4-174 x^3+81 x^2-16 x+1$\\
$Q_{9}^{c}$ & $48 x^9-6 x^8+50 x^7+63 x^6-172 x^5+79 x^4+30 x^3-39 x^2+12 x-1$\\
$Q_{9}^{d}$ & $30 x^9-108 x^8+63 x^7+3 x^6-29 x^5+87 x^4-155 x^3+97 x^2-21 x+1$\\
\hline
$Q_{10}^{a}$ & $171 x^{10}+315 x^9-1380 x^8+2340 x^7-2631 x^6+2176 x^5-1335 x^4+572 x^3-152 x^2+21 x-1$\\
$Q_{10}^{b}$ & $72 x^{10}+12 x^9+16 x^8-287 x^7+757 x^6-1025 x^5+883 x^4-537 x^3+219 x^2-51 x+5$\\
$Q_{10}^{c}$ & $270 x^{10}+762 x^9-1377 x^8+114 x^7+1426 x^6-1546 x^5+864 x^4-338 x^3+96 x^2-16 x+1$\\
$Q_{10}^{d}$ & $186 x^{10}-394 x^9+605 x^8-670 x^7+428 x^6+422 x^5-810 x^4+486 x^3-150 x^2+28 x-3$\\
\hline
$Q_{11}^{a}$ & $252 x^{11}+483 x^{10}-4456 x^9+7239 x^8-4462 x^7-226 x^6+1958 x^5-1294 x^4+482 x^3-121 x^2+18 x-1$\\
\hline
$Q_{12}^{a}$ & $1674 x^{12}+2766 x^{11}-9449 x^{10}+8532 x^9-2121 x^8-2040 x^7+2826 x^6-2908 x^5+2460 x^4-1302 x^3+383 x^2-56 x+3$\\
\hline
\end{tabular}
\caption{\label{TableOfIrreduciblePolynomials} Tables of $\mathbbm{Z}$-irreducible polynomials $Q_{n}^{\rm x}$ used in the definitions of polynomials $P_{n}^{\rm x}$ in Table \ref{TableOfBigPolynomials} .}
\end{table}
Consistency requires that in the limit $r\to 0$ the minimal model situation should be recovered. Indeed, in such a case one obtains:
\bea
w^{u*}_{10} &= & -2 Q_{3b} Q_{5}^{a} (x-1)^3,\;\;\;
w^{u*}_{\overline{126}}=-\sqrt{6}\frac{\alpha }{\eta} Q_{3}^{a} Q_{3}^{b} (x-1)^4 (3 x-1) ,\;\;\;
w^{u*}_{{126}}=\sqrt{6}\frac{\overline\alpha}{\eta } Q_{5}^{a} (x-1)^4 \left(2 x^2+x-1\right) ,\nn
\\
w^{u*}_{210}& = & -\frac{ \alpha  \overline\sigma}{\lambda  \omega }Q_{3b} Q_{3}^{c} (x-1)^4 x,\;\;\;
w^{u(1)*}_{120}=w^{u(2)*}_{120}=0,\nn
\\
w^{d*}_{10}&=&-2 Q_{3}^{b} Q_{5}^{a} (x-1)^3,\;\;\;
w^{d*}_{{126}}=-\sqrt{6}\frac{ \overline\alpha}{\eta } Q_{3}^{a} Q_{3}^{b} (x-1)^4 (3 x-1),\;\;\;
w^{d*}_{\overline{126}}=\sqrt{6}\frac{ \alpha }{\eta } Q_{5}^{a} (x-1)^4 \left(2 x^2+x-1\right),\nn 
\\
w^{d*}_{210}&=&-\frac{\overline\alpha\sigma}{\lambda  \omega }Q_{3}^{b}Q_{3}^{c} (x-1)^4 x  ,\;\;\;
w^{d(1)*}_{120}=w^{d(2)*}_{120}=0,
\nn
\label{weightsindecouplinglimit}
\eea
which is (up to an overall factor $-Q_{3}^{b}(x-1)^{4}$ and signs we shall comment on below) identical to the minimal scenario relations, c.f. formulae (C18), (C19) in \cite{Bajc:2004xe} in view of the comments given e.g. in \cite{Aulakh:2006vi}. We can also see that in such a case the $\bf 120_{\rm H}$ Higgs multiplet decouples as desired. 

\paragraph*{Note on convention and phase factors:} Notice that the phase convention used to derive formulae (\ref{doubletmassmatrix}), (\ref{upsectorweights}) and (\ref{downsectorweights}) has been chosen in such a way that the mass matrix is ``optically simple'', i.e. with positive multiples of the superpotential mass terms on the diagonal and without explicit $i$-factors in the 5th and 6th rows and columns.
This is also the reason why we have got some of the signs in (\ref{weightsindecouplinglimit}) different from those in the corresponding formulae in \cite{Bajc:2004xe} (notice in particular the extra minus sign at the 44 position of the relevant mass matrix in there).
As far as only the mass matrices are concerned, this has, of course, no physical significance and the physical spectra remain intact.
One must, however, pay attention to these effects upon getting to the interaction vertices, which are of course sensitive to the particular choice of phases in (\ref{doubletmassmatrix}) and (\ref{doubletweights}). This, in particular, is relevant for the matching of the ``microscopic theory'' (i.e. the $SO(10)$ model) onto the effective Yukawa sector sum-rules, that shall be studied thoroughly in the next section. 
\subsection{Matching NMSGUT to the Yukawa sector of the MSSM}
\subsubsection{Note on phases \& conventions}
Let us now focus on the form of the effective Yukawa sum-rules in the specific convention adopted in this study, which leads in particular to an ``optically simple'' form of (not only) the doublet mass matrix (\ref{doubletmassmatrix}). It is important to notice that the would-be simple-minded identification\footnote{The complex conjugation comes from the need to rewrite the defining basis fields (entering the Yukawa interactions in $W_{\rm Y}$) in terms of the mass eigenstates (that develop the electoroweak VEVs in the zero-mode direction) and the requirement of unitarity of transformation (\ref{doubletweights}).} $v^{u,d}_{10}=w^{u,d*}_{10}v_{u,d}$,  $v^{u,d}_{\overline{126}}=w^{u,d*}_{\overline{126}}v_{u,d}$, $v^{u,d(1)}_{120}=w^{u,d(1)*}_{120}v_{u,d}$ and $v^{u,d(2)}_{120}=w^{u,d(2)*}_{120}v_{u,d}$ is contrived because of its convention-dependence. Indeed, changing for instance the convention in such a way that the symbol $\Psi^{u}_{(1)}$ would be instead used for what we would call $-\Psi^{u}_{(1)}$ in the original convention (which, being a phase transformation, would be as good basis vector as the original one), the functional dependence of the corresponding $w_{120}^{u(1)*}$ weight factor on the ``microscopic'' parameters ($x$, $r$, $\alpha\ldots$) in the new convention would get an extra minus sign with respect to the form given in formula (\ref{upsectorweights}). Such a change, however, should lead to a change of the relative sign of terms in the square bracket in the first two equations in (\ref{NMSGUTsumrules}), which can not be absorbed into a redefinition of the Yukawa matrix\footnote{This, however, does not mean that there is no mapping between the entire parametric spaces of the original convention and the new one accounting for such a change, but (even if existed) it could be far from trivial to find. Furthermore, the problem gets even more serious when the change would be promoted to other sectors of the model.}. Apart from the phases, there could also be Clebsch-Gordon coefficients popping-up in the matching. Note also that this is a general issue of any matching between an effective and a ``microscopic'' theory. 

Therefore, in order for the information provided in this analysis to be self-contained and verifiable, the matching conditions should be carefully inspected and all the utilized conventions have to be fully specify. That is also why we have devoted a significant portion of Section (\ref{sect-massmatrices}) to the detailed specification of the relevant SM eigenvectors.

\subsubsection{Breaking  $SU(3)_{c}\otimes SU(2)_{L}\otimes U(1)_{Y}$ down to $SU(3)_{c}\otimes U(1)_{Q}$}

\paragraph{Doublet VEVs:}\mbox{}\\
Let us {\it define} the projections of the electroweak doublet VEVs onto the neutral components of the relevant defining basis doublets $H^{u,d}$, $\overline\Sigma^{u,d}$, $\Psi^{u,d}_{(1)}$ and $\Psi^{u,d}_{(2)}$ as follows:
\bea
\vev{H^{u}}\equiv u_{10}^{u},\;\; \vev{\overline \Sigma^{u}}\equiv u_{\overline{126}}^{u},\;\;\vev{\Psi^{u}_{(1)}}\equiv u_{120}^{u(1)},\;\;\vev{\Psi^{u}_{(2)}}\equiv u_{120}^{u(2)},\nn\\
\vev{H^{d}}\equiv u_{10}^{d},\;\;\vev{\overline \Sigma^{d}}\equiv u_{\overline{126}}^{d},\;\;\vev{\Psi^{d}_{(1)}}\equiv u_{120}^{d(1)},\;\;\vev{\Psi^{d}_{(2)}}\equiv u_{120}^{d(2)}.\;
\label{electroweakVEVsphysical}
\eea
The main virtue of this definition is that these factors {\it are} indeed simple functions of the decomposition weights in (\ref{doubletweights}) and the VEVs $v_{u}$ and $v_{d}$ of the MSSM light Higgs doublets ($\vev{h_{u,d}}\equiv v_{u,d}$), namely:
\bea\label{projectionsintermsofweights}
&& u^{u}_{10}=w^{u*}_{10}v_{u},\;\, u^{u}_{\overline{126}}=w^{u*}_{\overline{126}}v_{u},\;\, u^{u(1)}_{120}=w^{u(1)*}_{120}v_{u}\,\text{ and }\; u^{u(2)}_{120}=w^{u(2)*}_{120}v_{u}\nn \\
&& u^{d}_{10}=w^{d*}_{10}v_{d},\;\;  u^{d}_{\overline{126}}=w^{d*}_{\overline{126}}v_{d},\;\; u^{d(1)}_{120}=w^{d(1)*}_{120}v_{d}\;\text{ and }\; u^{d(2)}_{120}=w^{d(2)*}_{120}v_{d}
\eea
and thus, given the relevant MSSM doublet weight factors in formulae (\ref{upsectorweights}) and (\ref{downsectorweights}), can be readily computed from the underlying theory.

In order to be fully specific, let us also note that in our convention the electroweak doublet VEVs (\ref{electroweakVEVsphysical}) are spread over the following components of the defining $SO(10)$ tensors (\ref{components}): 
\bea
\vev{H_{3}}\!\!&=& \!\!\frac{i}{\sqrt{2}}(u_{10}^{d}-u_{10}^{u}),\;\;\vev{H_{4}}=\frac{1}{\sqrt{2}}(u_{10}^{d}+u_{10}^{u})\label{VEVprojections}
\nn\\
\vev{\overline \Sigma_{12356}}\!\!&=& \!\! \vev{\overline \Sigma_{12378}}= \vev{\overline \Sigma_{12390}}= \frac{i}{2\sqrt{3}}(u_{\overline{126}}^{d}-u_{\overline{126}}^{u}),\quad\;\;\;
\vev{\overline \Sigma_{12456}}= \vev{\overline \Sigma_{12478}}= \vev{\overline \Sigma_{12490}}=\frac{1}{2\sqrt{3}}(u_{\overline{126}}^{d}+u_{\overline{126}}^{u}),\nn\\
\vev{\overline \Sigma_{35678}}\!\!&=& \!\! \vev{\overline \Sigma_{35690}}= \vev{\overline \Sigma_{37890}}= -\frac{i}{2\sqrt{3}}(u_{\overline{126}}^{d}+u_{\overline{126}}^{u}),\;\;\;\;
\vev{\overline \Sigma_{45678}}= \vev{\overline \Sigma_{45690}}= \vev{\overline \Sigma_{47890}}=-\frac{1}{2\sqrt{3}}(u_{\overline{126}}^{d}-u_{\overline{126}}^{u}),\nn\\
\vev{\Psi_{123}}\!\!&=& \!\!\frac{1}{\sqrt{2}}(u_{120}^{d(1)}-u_{120}^{u(1)}),\;\;
\vev{\Psi_{124}}= -\frac{i}{\sqrt{2}}(u_{120}^{d(1)}+u_{120}^{u(1)}),\label{ewVEVsoncomponents}\\
\vev{\Psi_{356}}\!\!&=& \!\!
\vev{\Psi_{378}}= 
\vev{\Psi_{390}}= -\frac{1}{\sqrt{6}}(u_{120}^{d(2)}-u_{120}^{u(2)}),\qquad\;\;\;\;
\vev{\Psi_{456}}=
\vev{\Psi_{478}}= 
\vev{\Psi_{490}}= \frac{i}{\sqrt{6}}(u_{120}^{d(2)}+u_{120}^{u(2)}),\nn
\eea
This information shall be used later on in Section \ref{sect-matching} when it comes to the derivation of the matching conditions for the effective sum-rule VEV factors  $v_{\bf R}^{u,d}$ in (\ref{NMSGUTsumrules}) in terms of the $u_{\bf R}^{u,d}$ symbols fixing our convention for the neutral components of the defining basis $SU(2)_{L}$ doublets (\ref{electroweakVEVsphysical}). 
\vskip 5mm
\paragraph{Induced $SU(2)_{L}$-triplet VEV:}\mbox{}\\
Apart from the $SU(2)_{L}$ doublet VEVs, there is a pair of VEVs relevant for the Majorana sector of the model, namely those of the colourless $SU(2)_{L}$ singlet $\Delta_{R}^{0}$ corresponding to the neutral component of the $SU(2)_{R}$ triplet $\Delta_{R}=(10,1,3)_{\overline{126}}$ (this VEV is actually responsible for the $SU(2)_{R}\otimes U(1)_{B-L}\to U(1)_{Y}$ breakdown) and the colourless $SU(2)_{L}$ triplet $\Delta_{L}^{0}$ residing in $\Delta_{L}=(\overline{10},3,1)_{\overline{126}}$. Denoting the former by
\be\label{singletVEV}
\vev{\Delta_{R}^{0}}\equiv U_{R}
\ee
(which is just a more ``physical'' notation for $\overline\sigma$; indeed, as one can see from (\ref{VEVs}) these symbols are equivalent and we use the latter only for sake of clarity\footnote{The capital letter is used to express the fact that unlike the triplet VEV $u_{L}$ constrained by the $\rho$-parameter of the SM to be well within 1 GeV region, the $B-L$ breaking VEV of the $SU(2)_{R}$ triplet is typically around the GUT scale.}). Vanishing of the GUT-scale $F$-terms then requires a non-zero VEV to be induced on\footnote{The convenience of the extra $i$ in the definition (\ref{tripletVEV}) shall become obvious from what follows.} $\Delta_{L}^{0}$:
\be\label{tripletVEV}
\vev{\Delta_{L}^{0}}\equiv iu_{L}.
\ee
In particular, the relevant formula reads (c.f. for example \cite{Bajc:2005qe}) :
\be\label{inducedtripletVEV}
u_{L}=\frac{v_{210}^{u}}{m_{\Sigma}+\eta(3a-p)}\left[
\left(\alpha u_{10}^{u}-\sqrt{6}\eta u_{\overline{126}}^{u}\right)+
2\beta\left(v_{120}^{u(1)}+\sqrt{3}v_{120}^{u(2)}\right)
\right]
\ee  
where, as in the case of the other doublets (\ref{projectionsintermsofweights}), $v_{210}^{u}=w_{210}^{u*}v_{u}$ and the relevant weight factor is given in formula (\ref{upsectorweights}).
Note also that if $\bf 120_{\rm H}$ decouples (i.e. for $\beta\to 0$) this formula is reduced (up to the minus sign in the second term corresponding to a different convention being used) to the minimal model result given in \cite{Bajc:2005qe}.  It is remarkable that both factors, i.e. the minimal model contribution 
$\propto \alpha u_{10}^{u}-\sqrt{6}\eta u_{\overline{126}}^{u}$ as well as the extra piece due to $\bf 120_{\rm H}$ proportional to $v_{120}^{u(1)}+\sqrt{3}v_{120}^{u(2)}$ share a common feature that the Clebsch-Gordon coefficients therein exactly cancel the overall constants in the relevant weight factors and the resulting polynomials admit for a great further simplification, which can be viewed as an indication of consistency of our results. 

As before, to be fully explicit, let us remark that the $B-L$ breaking VEVs $U_{R}$ and $u_{L}$ from (\ref{singletVEV}) and (\ref{tripletVEV}) are spread over the following components of the relevant $SO(10)$ tensors:
\bea
-\vev{\overline \Sigma_{12379}}\!\!&=& \!\! \vev{\overline \Sigma_{123810}}\!=\! \vev{\overline \Sigma_{13670}}\!=\! \vev{\overline \Sigma_{13689}}\!=\! i\vev{\overline \Sigma_{14579}}\!=\!-i\vev{\overline \Sigma_{14580}}\!=\!-i\vev{\overline \Sigma_{14670}}\!=\!-i\vev{\overline \Sigma_{14689}}
\!=\!\frac{(u_{L}+U_{R}),}{4\sqrt{2}}\nn\\
-\vev{\overline \Sigma_{23570}}\!\!&=& \!\! -\vev{\overline \Sigma_{23589}}\!=\! -\vev{\overline \Sigma_{23679}}\!=\! \vev{\overline \Sigma_{23680}}\!=\! i\vev{\overline \Sigma_{24570}}\!=\!i\vev{\overline \Sigma_{24589}}\!=\!i\vev{\overline \Sigma_{24679}}\!=\!-i\vev{\overline \Sigma_{24680}}
\!=\!\frac{(u_{L}+U_{R})}{4\sqrt{2}},\nn\\
i\vev{\overline \Sigma_{13570}}\!\!&=& \!\! i\vev{\overline \Sigma_{13589}}\!=\! i\vev{\overline \Sigma_{13679}}\!=\! -i\vev{\overline \Sigma_{13680}}\!=\! \vev{\overline \Sigma_{14570}}\!=\!\vev{\overline \Sigma_{14589}}\!=\!\vev{\overline \Sigma_{14679}}\!=\!-\vev{\overline \Sigma_{14680}}
\!=\!\frac{(u_{L}-U_{R})}{4\sqrt{2}},\nn\\
-i\vev{\overline \Sigma_{23579}}\!\!&=& \!\! i\vev{\overline \Sigma_{23580}}\!=\! i\vev{\overline \Sigma_{23670}}\!=\! i\vev{\overline \Sigma_{23689}}\!=\! -\vev{\overline \Sigma_{24579}}\!=\!\vev{\overline \Sigma_{24580}}\!=\!\vev{\overline \Sigma_{24670}}\!=\!\vev{\overline \Sigma_{24689}}
\!=\!\frac{(u_{L}-U_{R})}{4\sqrt{2}}.
\label{ewVEVsoncomponentsMajorana}
\eea
\subsubsection{Matching NMSGUT to the effective Yukawa sum-rules\label{sect-matching}} 
The rest of this section will be devoted to a derivation of the effective sum-rules of the form (\ref{NMSGUTsumrules}) in terms of these VEVs and the identification of the ``effective'' Yukawa couplings $Y_{10}$, $Y_{126}$ and $Y_{120}$ in terms of the superpotential couplings $f_{10}$, $f_{\overline{126}}$ and $f_{120}$ in $W_{\rm Y}$, c.f. formula (\ref{extendedYukawagrouptheory}).

In what follows we shall use the method \cite{Mohapatra:1979nn} to work out the three relevant structures. In particular, we shall be using the embedding of the $SO(10)$ spinorial matter multiplets into the $\psi_{+}$ sector of  \cite{Mohapatra:1979nn}. Note that our convention is such that the ``colourless'' indices $\{1,2,3,4\}$ used in this study correspond to the $\{7,8,9,10\}$ sector of Mohapatra and Sakita and similarly our $\{5,6,7,8,9,10\}$ indices can be identified with their $\{2,1,4,3,6,5\}$ sector. With this at hand one can write the effective MSSM matter mass terms in a generic form (suppressing the family indices):
\be\label{SakitaStructure}
{\cal L}_{\mathrm m}=
f_{10}\langle \psi_{+}^{*} |B\, \Gamma_{i}|\psi_{+}\rangle \vev{H_{i}}+
\frac{1}{5!}f_{\overline{126}}\langle \psi_{+}^{*} |B\, \Gamma_{i}\Gamma_{j}\Gamma_{k}\Gamma_{l}\Gamma_{m}|\psi_{+}\rangle \vev{\overline\Sigma_{ijklm}}+
\frac{1}{3!}f_{\overline{120}}\langle \psi_{+}^{*} |B\, \Gamma_{i}\Gamma_{j}\Gamma_{k}|\psi_{+}\rangle \vev{\Psi_{ijk}}
\ee
where 
the $B$ and $\Gamma_{i}$ matrices as well as the bra's $\langle \psi_{+}^{*}|$ and ket's $|\psi_{+}\rangle$  are defined in \cite{Mohapatra:1979nn}. 
\vskip2mm
\paragraph{$SU(2)_{L}$-doublet VEVs \& Dirac mass sum-rules:}\mbox{}\\
Working out the structure (\ref{SakitaStructure}) together with (\ref{ewVEVsoncomponents}) one first obtains the effective Dirac mass matrices in the form:
\bea
Y_{u}v_{u}&= &
\sqrt{2}if_{10}u_{10}^{u}-\frac{2}{\sqrt{3}}if_{\overline{126}}u_{\overline{126}}^{u}+\sqrt{2}if_{120}\left(u_{120}^{u(1)}-\frac{1}{\sqrt{3}}u_{120}^{u(2)}\right) 
\nn\\
Y_{\nu}v_{u}&= & 
\sqrt{2}if_{10}u_{10}^{u}+3\frac{2}{\sqrt{3}}if_{\overline{126}}u_{\overline{126}}^{u}+\sqrt{2}if_{120}\left(u_{120}^{u(1)}+{\sqrt{3}}u_{120}^{u(2)}\right) 
\nn\\
Y_{d}v_{d}&=& 
\sqrt{2}if_{10}u_{10}^{d}+\frac{2}{\sqrt{3}}if_{\overline{126}}u_{\overline{126}}^{d}-\sqrt{2}if_{120}\left(u_{120}^{d(1)}+\frac{1}{\sqrt{3}}u_{120}^{d(2)}\right) 
\label{sumrulesrawform}\\
Y_{e}v_{d}&= & 
\sqrt{2}if_{10}u_{10}^{d}-3\frac{2}{\sqrt{3}}if_{\overline{126}}u_{\overline{126}}^{d}-\sqrt{2}if_{120}\left(u_{120}^{d(1)}-{\sqrt{3}}u_{120}^{d(2)}\right). 
\nn
\eea
Denoting $Y_{10}\equiv \sqrt{2}if_{10}$, $Y_{126}\equiv 2if_{\overline{126}}/{\sqrt{3}}$ and  $Y_{120}\equiv \sqrt{2}if_{120}$, the effective sum-rules in the notation (\ref{NMSGUTsumrules}) can be then recovered provided:
\bea
v_{10}^{u}\equiv u_{10}^{u},\;\;
v_{\overline{126}}^{u}\equiv -u_{\overline{126}}^{u},\;\;
v_{120}^{u(1)}\equiv u_{120}^{u(1)},\;\;
v_{120}^{u(2)}\equiv -\frac{1}{\sqrt{3}}u_{120}^{u(2)},\;\;\nn\\
v_{10}^{d}\equiv u_{10}^{d},\;\;
v_{\overline{126}}^{d}\equiv u_{\overline{126}}^{d},\;\;
v_{120}^{d(1)}\equiv -u_{120}^{d(1)},\;\;
v_{120}^{d(2)}\equiv -\frac{1}{\sqrt{3}}u_{120}^{d(2)},\;\;\label{matching}
\eea
which, together with formula (\ref{projectionsintermsofweights}) yields the desired matching between the weight factors given by formulae (\ref{upsectorweights}) and (\ref{downsectorweights}) and the effective electroweak weight factors in (\ref{NMSGUTsumrules}) in our convention. 

\vskip5mm
\paragraph{$SU(2)_{L}$-singlet and triplet VEVs \& Majorana masses:}\mbox{}\\
For sake of completeness let us present also the relevant Majorana sector matching formulae. After some tedium, equations (\ref{ewVEVsoncomponentsMajorana}) and (\ref{SakitaStructure}) yield:
 \be
 {\cal L}_{\mathrm m}^{\mathrm M}= \frac{1}{2}(\nu^{c}_{L})^{T}C^{-1}M_{\nu}^{R}\nu^{c}_{L}+ \frac{1}{2}(\nu_{L})^{T}C^{-1}M_{\nu}^{L}\nu_{L}+h.c.\quad \text{with}\quad 
 M_{\nu}^{R}=-4\sqrt{2}if_{\overline{126}}U_{R}\;\text{ and }\;   M_{\nu}^{L}=4\sqrt{2}if_{\overline{126}}u_{L}.
 \ee
 Note that due to the properties of $f_{\overline{126}}$ the Majorana masses are indeed symmetric in the family space, as desired. 
 Restoring the effective Yukawa couplings, the matching conditions for the effective VEV factors in the Majorana sector of (\ref{NMSGUTsumrules}) read:
 \be\label{matchingMajorana}
V_{R}=-2\sqrt{6}\,U_{R}\;\text{ and }\;   v_{L}=2\sqrt{6}\,u_{L}.
 \ee
 Note that these results correspond to those obtained previously in \cite{Bajc:2005qe} combined with \cite{Aulakh:2004hm,Aulakh:2005bd} and thus provide a further nontrivial consistency check of our calculation. 
\subsection{Final remarks} 
With all these results at hand, all the ingredients necessary for fitting the effective sum-rules (\ref{NMSGUTsumrules}) in the framework of the full-featured NMSGUT have been discussed. Indeed, matching conditions (\ref{matching}) and (\ref{matchingMajorana}), when supplemented with the prescriptions (\ref{singletVEV})  and (\ref{inducedtripletVEV}) together with the translation tables  (\ref{SMvacuumparameters}) and (\ref{projectionsintermsofweights}) and the explicit formulae for the weight factors (\ref{upsectorweights}) and (\ref{downsectorweights}) can be used to rewrite all the weight factors in (\ref{NMSGUTsumrules}) in terms of the ``microscopic'' parameters $\{x, m_{\Phi}, r, \alpha,\overline{\alpha},\eta,\lambda,\beta, \overline{\beta},\gamma,\rho\}$.

\section{Conclusions}
It has been shown several years ago that the renormalizable SUSY $SO(10)$ grand-unified model with the simplest potentially realistic Higgs sector spanned over the representations ${\bf 10}_{\rm H}$, ${\bf \overline{126}}_{\rm H}\oplus {\bf 126}_{\rm H}$ and ${\bf 210}_{\rm H}$ experiences severe difficulties in accommodating the low-energy Yukawa sector constraints stemming from the observed patterns of the quark and lepton masses and mixing parameters.  
It was argued that the troubles emerge due to the antagonism between the need for a $B-L$ breaking scale to be slightly suppressed relative to the GUT scale (which is vital in order to bring the type-I seesaw contribution to the neutrino masses into play) on one side and the significant difference between the second generation quark and lepton masses calling for essentially the opposite on the other.

One of the most popular simple extensions of the minimal framework that could in principle resolve this issue consists in employing an extra 120-dimensional three-index antisymmetric tensor representation in the Higgs sector providing for a new contribution to the effective Yukawa sector sum-rules, thus relaxing the tight link between the seesaw scale and the second generation hierarchy.

Due to the rather complicated structure of the Higgs sector, the key ingredient of any quantitative analysis of such kind of models is a thorough understanding of the relevant Higgs spectra and the corresponding Higgs mixing patterns. It is remarkable that even in the case of the minimal model (which was formulated at the beginning of 1980's) the mathematical complexity of the 472-dimensional Higgs sector did not admit for drawing reliable statements about the viability of the theory until a couple of  years ago when the complete analysis of the ${\bf 10}_{\rm H}$, ${\bf \overline{126}}_{\rm H}\oplus {\bf 126}_{\rm H}$ and ${\bf 210}_{\rm H}$ Higgs sector was first published \cite{Bajc:2004xe}.

In the current paper we provided a very detailed and maximally self-contained analysis of the 592-dimensional Higgs potential of the next-to minimal SUSY SO(10) model (consisting of ${\bf 10}_{\rm H}$, ${\bf \overline{126}}_{\rm H}\oplus {\bf 126}_{\rm H}$, ${\bf 210}_{\rm H}$ and ${\bf 120}_{\rm H}$ multiplets) focusing on the effects of the extra ${\bf 120}_{\rm H}$ in the game. Since there are no extra Standard Model singlets in ${\bf 120}_{\rm H}$ the symmetry breaking pattern of the minimal model remains intact which, in turn, simplifies the analysis considerably and admits for adopting the useful notation of \cite{Bajc:2004xe} for the case under consideration.
 
In particular, all the GUT-scale Higgs sector mass matrices have been written in detail together with a thorough description of the basis states and the relevant conventions in each of the different SM sectors. Focusing subsequently on the sum-rules for the effective Yukawa sector emerging under the GUT-scale, and in particular to the masses of the matter fermions, the matching of all the relevant effective building blocks to the microscopic structure of the model was investigated. A set of nontrivial consistency checks was also provided: 1) in the limit of a decoupling ${\bf 120}_{\rm H}$ all the mass matrices given in this study reduce to the minimal models structures of \cite{Bajc:2004xe} (up to unphysical rearrangements) and 2) the 33 Goldstone modes were shown to be present in the Higgs sector spectra so that the proper Standard Model gauge structure emerges at low energies.
\section*{Note added}
A day before finishing ver.1 of this manuscript the author's attention was drawn to the preprint \cite{Aulakh:2006hs} where the relevant part of the next-to minimal SUSY $SO(10)$ model has been previously studied from a similar perspective. As far as one can see through the jungle of different notation, normalization and conventions the results therein agree with those given in this study. Moreover, since the method we employed is different I believe that the current analysis is worth and does indeed provide a valuable and an independent survey of many of the crucial and technically rather demanding prerequisites of any numerical analysis of the NMSGUT scenario. Despite from that, there could still be a good case for even a further check of ours as well as Aulakh \& Garg's results, in particular when it comes to the phases and matching(s).

Apart from this, the current study is rather detailed on various aspects and we aimed onto making it maximally self-contained so that a careful and patient reader could potentially reproduce all the results with just the ingredients given here and in the 'canonical' MSGUT reference \cite{Bajc:2004xe}. On top of that, some of the extra information provided in Sections \ref{sect-massmatrices} and \ref{sect-yukawasector} and in particular in Sections \ref{goldstones} and in Appendix \ref{decompositions} does not have any counterpart in \cite{Aulakh:2006hs}, due to a different method used therein.
\section*{Acknowledgments}
I am very grateful to Stefano Bertolini, Thomas Schwetz and Takeshi Fukuyama for discussions throughout preparing this manuscript. I am indebted to Charanjit Aulakh for pointing out the notation inconsistency in ver.1 of the ArXiv posting. The work was supported by the PPARC Rolling Grant PPA/G/S/2003/00096.
\appendix
\section{$SU(3)_{c}\otimes SU(2)_{L}\otimes U(1)_{Y}$ components of the $SO(10)$ tensors \label{decompositions}}
\subsection{$SU(3)_{c}\otimes SU(2)_{L}\otimes U(1)_{Y}$ content of $\bf 10$, $\bf 126\oplus \overline{126}$ and  $\bf 210$ of $SO(10)$}
The mapping of these $SO(10)$ representations, in particular their submultiplets with definite Standard Model quantum numbers onto the defining bases $H_{i}$, $\Sigma_{ijklm}$, $\overline{\Sigma}_{ijklm}$ and $\Phi_{ijkl}$ has been given in detail previously in the work \cite{Bajc:2004xe}. As an independent cross-check of these results we have repeated the analysis and our results entirely confirm those obtained in \cite{Bajc:2004xe}, up to different phase factors. These are, however, irrelevant as far as the decompositions are concerned\footnote{This is true unless the states under consideration belong to the same SM multiplet -- in such a case their relative phase is fixed by the phase structure of the relevant simple roots.}, but crucial once it comes to the interaction vertices. Thus, we shall not repeat all the lengthy prescriptions for relevant maps here, but rather comment on how an interested reader can translate the tables IV, V, VI and VII of \cite{Bajc:2004xe} to conform the conventions used in this study.
\begin{itemize}
\item{First, the generic symbols $[i,j,k,l]$ in tables IV and V of  \cite{Bajc:2004xe} correspond to $\Phi[i,j,k,l]$ in our notation provided  
\be
\label{Phiantisymmetrized}\Phi[i,j,k,l]\equiv\frac{1}{\sqrt{4!}}\left(\Phi_{ijkl}-\Phi_{ijlk}+\Phi_{iljk}-\Phi_{lijk}+\ldots\right);\ee
similarly $[i,j,k,l,m]$ of table VII of  \cite{Bajc:2004xe} stand for the totally antisymmetrized combinations of the defining basis vectors of $\bf 126$, i.e.:
\be
\Sigma[i,j,k,l,m]\equiv\frac{1}{\sqrt{5!}}\left(\Sigma_{ijklm}-\Sigma_{ijkml}+\Sigma_{ijmkl}-\Sigma_{imjkl}+\ldots\right)
\ee
and $[i]$ in table VI of \cite{Bajc:2004xe} is our $H[i]\equiv H_{i}$.
}
\item{The last columns  of the tables IV, V, VI and VII in \cite{Bajc:2004xe} contain the (parts of the) maps of the SM states with the quantum numbers specified in the respective first two columns (which can be easily translated to our $(C^{c}, T^{c}_{3},  T^{c}_{8}, B-L, C^{L},T^{L}_{3}, C^{R}, T^{R}_{3})$ notation) in terms of the basic antisymmetric structures identified above.}
\item{The overall phase of this map (i.e. the overall phase of the coefficients in the third column of the relevant tables in \cite{Bajc:2004xe}) is unphysical unless the interactions (and mass matrices ) are inspected. Our phase convention used upon deriving the mass matrices in this study is then specified by providing the phase factor of the first chunk of the map in the tables specifying the quantum numbers of the row and column states throughout Section \ref{sect-massmatrices}.}
\item{The overall normalization factor is not displayed, but can be readily computed.}
\end{itemize}
As an example, let us consider the  $(0,0,0,+2,\tfrac{3}{4},-\tfrac{1}{2},\tfrac{3}{4},-\tfrac{1}{2})_{210}$ submultiplet of $({10},2,2)_{210}$ contributing to the MSSM-like doublets with the mass matrix (\ref{doubletmassmatrix}). This state corresponds to the first item of the $(\overline{10},2,2)$ section of table V in \cite{Bajc:2004xe}. Applying the dictionary above and fixing the overall phase from the relevant table under the mass matrix (\ref{doubletmassmatrix}) we get for instance:
\bea
& & (0,0,0,+2,\tfrac{3}{4},-\tfrac{1}{2},\tfrac{3}{4},-\tfrac{1}{2})_{210}\equiv 
\frac{1}{4}\times \nn \\
& & (-i \Phi [1,5,7,9]\!+\!\Phi [1,5,7,10]\!+\!\Phi [1,5,8,9]\!+i \Phi
   [1,5,8,10]\!+\!\Phi [1,6,7,9]\!+i \Phi [1,6,7,10]\!+i \Phi [1,6,8,9]\!-\!\Phi
   [1,6,8,10]\!\nn\\
& &-\!\Phi [2,5,7,9]\!-i \Phi [2,5,7,10]\!-i \Phi [2,5,8,9]\!+\!\Phi
   [2,5,8,10]\!-i \Phi [2,6,7,9]\!+\!\Phi [2,6,7,10]\!+\!\Phi [2,6,8,9]\!+i \Phi
   [2,6,8,10])\nn
\eea
which is the correct result\footnote{For obvious reasons, we do not further expand the antisymmetrized combinations along the lines of (\ref{Phiantisymmetrized}), which in the case under consideration would account not for 16 but for 1920 terms in total.} conforming our phase and notation conventions. In a similar manner one can construct all the remaining maps\footnote{Note that, unfortunately, there seems to be a typo in table V of \cite{Bajc:2004xe} in the $T_{3}^{R}=\pm$ assignment of the $(10,2,2)$ sector. Our results are such that all theses signs should be opposite (which can be further justified by the shape of the SO(10) invariants).}. 
\subsection{$SU(3)_{c}\otimes SU(2)_{L}\otimes U(1)_{Y}$ content of $\bf 120$ of $SO(10)$}
The SM decomposition and the mapping of bases for $\bf 120$ of $SO(10)$ has not been provided in  \cite{Bajc:2004xe} though. Thus, in this section we shall give explicit decompositions of all the submultiplets of ${\bf 120}_{\rm H}$ that have been used in the text as a basis for the various mass matrices. We shall give just a representative of each of the sectors - an interested reader can obtain all the other weights within the multiplet under consideration from the simple roots of the relevant Lie-algebras.  

As before, we shall use the $\Psi[i,j,k]$ symbols to represent the properly normalized totally antisymmetric combination of the defining components of $\bf 120$, i.e. 
\be\Psi[i,j,k]\equiv \frac{1}{\sqrt{6}}(\Psi_{ijk}-\Psi_{ikj}+\Psi_{jki}-\Psi_{jik}+\Psi_{kij}-\Psi_{kji})\ee
The results are given in Table \ref{120table}. The phase convention used therein is the same as the one used to derive the mass matrices in section \ref{sect-massmatrices}.
\begin{table}[h]
\begin{tabular}{|c|c|c|}
\hline 
Pati-Salam origin & ($C^{c}\!\!, T^{c}_{3},  T^{c}_{8}, B-L, C^{L}\!\!,T^{L}_{3}, C^{R}\!\!,  T^{R}_{3}$) & mapping onto the (antisymmetrized) defining basis\\
\hline
$(\overline{10},1,1)_{120}$ & $(0,0,0,-2,0,0,0,0)_{120}$ & 
$-\Psi [5,7,9]-i \Psi [5,7,10]-i \Psi[5,8,9]+\Psi [5,8,10]$\\
& & $-i \Psi [6,7,9]+\Psi[6,7,10]+\Psi [6,8,9]+i \Psi [6,8,10]$\\
$(10,1,1)_{120}$ & $(0,0,0,+2,0,0,0,0)_{120}$ & 
$-\Psi [5,7,9]+i \Psi [5,7,10]+i \Psi[5,8,9]+\Psi [5,8,10]$\\
& & $+i \Psi [6,7,9]+\Psi[6,7,10]+\Psi [6,8,9]-i \Psi [6,8,10]$\\
\hline
$(1,2,2)_{120}$ & $(0,0,0,0,\tfrac{3}{4},+\tfrac{1}{2},\tfrac{3}{4},-\tfrac{1}{2})_{120}$ & $-\Psi [1,2,3]-i \Psi[1,2,4]$ \\
$(15,2,2)_{120}$ & $(0,0,0,0,\tfrac{3}{4},+\tfrac{1}{2},\tfrac{3}{4},-\tfrac{1}{2})_{120}$ &
 $-\Psi [3,5,6]-\Psi [3,7,8]-\Psi [3,9,10]$ \\
 & & $i\Psi [4,5,6]-i \Psi [4,7,8]-i \Psi[4,9,10]$ \\
 $(1,2,2)_{120}$ & $(0,0,0,0,\tfrac{3}{4},-\tfrac{1}{2},\tfrac{3}{4},+\tfrac{1}{2})_{120}$ & $-\Psi [1,2,3]+i \Psi[1,2,4]$ \\
 $(15,2,2)_{120}$ & $(0,0,0,0,\tfrac{3}{4},-\tfrac{1}{2},\tfrac{3}{4},+\tfrac{1}{2})_{120}$ & 
 $-\Psi [3,5,6]-\Psi [3,7,8]-\Psi [3,9,10]$\\
 & & $+i\Psi [4,5,6]+i \Psi [4,7,8]+i \Psi[4,9,10]$ \\
\hline
$(6,1,3)_{120}$ & $(\frac{4}{3},0,+\frac{1}{\sqrt{3}},+\frac{2}{3},0,0,2,-1)_{{120}}$ & 
$-i \Psi [1,3,9]-\Psi [1,3,10]+\Psi[1,4,9]-i \Psi [1,4,10]$ \\
& & $+\Psi [2,3,9]-i\Psi [2,3,10]+i \Psi [2,4,9]+\Psi[2,4,10]$ \\
 $(6,1,3)_{120}$ & $(\frac{4}{3},0,-\frac{1}{\sqrt{3}},-\frac{2}{3},0,0,2,+1)_{{120}}$ & 
 $i \Psi [1,3,9]-\Psi [1,3,10]+\Psi[1,4,9]+i \Psi [1,4,10]$ \\
 & & $+\Psi [2,3,9]+i\Psi [2,3,10]-i \Psi [2,4,9]+\Psi[2,4,10]$ \\
 $(10,1,1)_{120}$ & $(\frac{4}{3},0,+\frac{1}{\sqrt{3}},+\frac{2}{3},0,0,0,0)_{{120}}$ & $-\Psi [5,6,9]+i \Psi [5,6,10]-\Psi [7,8,9]+i
   \Psi [7,8,10]$ \\
$(6,1,3)_{120}$ & $(\frac{4}{3},0,+\frac{1}{\sqrt{3}},+\frac{2}{3},0,0,2,0)_{{120}}$ & $-\Psi [1,2,9]+i \Psi [1,2,10]-\Psi [3,4,9]+i
   \Psi [3,4,10]$ \\
$(\overline{10},1,1)_{120}$ & $(\frac{4}{3},0,-\frac{1}{\sqrt{3}},-\frac{2}{3},0,0,0,0)_{{120}}$ & $-\Psi [5,6,9]-i \Psi
   [5,6,10]-\Psi [7,8,9]-i \Psi [7,8,10]$ \\
$(6,1,3)_{120}$ & $(\frac{4}{3},0,-\frac{1}{\sqrt{3}},-\frac{2}{3},0,0,2,0)_{{120}}$ & $-\Psi [1,2,9]-i \Psi
   [1,2,10]-\Psi [3,4,9]-i \Psi [3,4,10]$ \\
$(6,1,3)_{120}$ & $(\frac{4}{3},0,+\frac{1}{\sqrt{3}},+\frac{2}{3},0,0,2,+1)_{{120}}$ & 
$ +i \Psi [1,3,9]+\Psi [1,3,10]+\Psi [1,4,9]-i \Psi [1,4,10]$ \\
& & $+\Psi [2,3,9]-i \Psi [2,3,10]-i \Psi [2,4,9]-\Psi [2,4,10]$ \\
$(6,1,3)_{120}$ & $(\frac{4}{3},0,-\frac{1}{\sqrt{3}},-\frac{2}{3},0,0,2,-1)_{{120}}$ & 
$-i \Psi [1,3,9]+\Psi [1,3,10]+\Psi [1,4,9]+i \Psi [1,4,10]$ \\
& & $+\Psi [2,3,9]+i \Psi [2,3,10]+i \Psi [2,4,9]-\Psi [2,4,10]$ \\
\hline
 $(15,2,2)_{120}$ & $(\frac{4}{3},0,+\frac{1}{\sqrt{3}},-\frac{4}{3},\frac{3}{4},+\frac{1}{2},\frac{3}{4},-\frac{1}{2})_{{120}}$ & 
 $ -\Psi [3,5,7]-i \Psi [3,5,8]-i \Psi [3,6,7]+\Psi [3,6,8]$ \\
& & $-i\Psi [4,5,7]+\Psi [4,5,8]+\Psi [4,6,7]+i \Psi [4,6,8]$ \\
$(15,2,2)_{120}$ & $(\frac{4}{3},0,-\frac{1}{\sqrt{3}},+\frac{4}{3},\frac{3}{4},-\frac{1}{2},\frac{3}{4},+\frac{1}{2})_{{120}}$ & 
$-\Psi [3,5,7]+i \Psi [3,5,8]+i \Psi [3,6,7]+\Psi [3,6,8]$ \\
& & $+i\Psi [4,5,7]+\Psi [4,5,8]+\Psi [4,6,7]-i \Psi [4,6,8]$ \\
$(15,2,2)_{120}$ & $(\frac{4}{3},0,+\frac{1}{\sqrt{3}},-\frac{4}{3},\frac{3}{4},+\frac{1}{2},\frac{3}{4},+\frac{1}{2})_{{120}}$ & 
$-i \Psi [1,5,7]+\Psi [1,5,8]+\Psi [1,6,7]+i \Psi [1,6,8]$ \\
& & $-\Psi[2,5,7]-i \Psi [2,5,8]-i \Psi [2,6,7]+\Psi [2,6,8]$ \\
$(15,2,2)_{120}$ & $(\frac{4}{3},0,-\frac{1}{\sqrt{3}},+\frac{4}{3},\frac{3}{4},-\frac{1}{2},\frac{3}{4},-\frac{1}{2})_{{120}}$ & 
$ i \Psi [1,5,7]+\Psi [1,5,8]+\Psi [1,6,7]-i \Psi [1,6,8]$ \\
& & $-\Psi[2,5,7]+i \Psi [2,5,8]+i \Psi [2,6,7]+\Psi [2,6,8]$ \\
\hline
$(6,3,1)_{120}$ & $(\frac{4}{3},0,+\frac{1}{\sqrt{3}},+\frac{2}{3},2,0,0,0)_{{120}}$ & $+i \Psi [1,2,9]+\Psi [1,2,10]-i \Psi [3,4,9]-\Psi [3,4,10]$ \\
$(6,3,1)_{120}$ & $(\frac{4}{3},0,-\frac{1}{\sqrt{3}},-\frac{2}{3},2,0,0,0)_{{120}}$ & $-i \Psi [1,2,9]+\Psi [1,2,10]+i \Psi [3,4,9]-\Psi[3,4,10]$ \\
\hline
$(15,2,2)_{120}$ & $(3,-1,0,0,\frac{3}{4},-\frac{1}{2},\frac{3}{4},-\frac{1}{2})_{{120}}$ & 
$\Psi [1,5,7]+i \Psi [1,5,8]-i \Psi[1,6,7]+\Psi [1,6,8]$ \\
& & $+i \Psi [2,5,7]-\Psi[2,5,8]+\Psi [2,6,7]+i \Psi [2,6,8]$ \\
$(15,2,2)_{120}$ & $(3,+1,0,0,\frac{3}{4},+\frac{1}{2},\frac{3}{4},+\frac{1}{2})_{{120}}$ & 
$\Psi [1,5,7]-i \Psi [1,5,8]+i \Psi[1,6,7]+\Psi [1,6,8]$ \\
& & $-i \Psi [2,5,7]-\Psi[2,5,8]+\Psi [2,6,7]-i \Psi [2,6,8]$ \\
\hline
$(10,1,1)_{120}$ & $(\frac{10}{3},0,+\frac{2}{\sqrt{3}},-\frac{2}{3},0,0,0,0)_{{120}}$ & 
$\Psi [5,7,9]-i \Psi [5,7,10]+i \Psi[5,8,9]+\Psi [5,8,10]$ \\
& & $+i \Psi [6,7,9]+\Psi[6,7,10]-\Psi [6,8,9]+i \Psi [6,8,10]$ \\
$(\overline{10},1,1)_{120}$ & $(\frac{10}{3},0,-\frac{2}{\sqrt{3}},+\frac{2}{3},0,0,0,0)_{{120}}$ & 
$\Psi [5,7,9]+i \Psi [5,7,10]-i \Psi[5,8,9]+\Psi [5,8,10]$ \\
& & $-i \Psi [6,7,9]+\Psi[6,7,10]-\Psi [6,8,9]-i \Psi [6,8,10]$ \\
\hline 
\end{tabular}
\caption{\label{120table}Mapping of the SM submultiplets of the three-index antisymmetric tensor of $SO(10)$ onto the defining basis vectors. }
\end{table}


\end{document}